\documentclass[twocolumn]{aastex631}
\usepackage{ulem}
\usepackage{comment}
\usepackage[shadow]{todonotes}	
\usepackage{color, soul}	

\newcommand{\eg}[1]{(e.g. \citealt{#1})}

\shortauthors{
Wong, Steiner et~al.
}

\shorttitle{
Hard X-ray Emission of 3C~264
}

\begin{document}
\title{\textit{NuSTAR} Observation of the TeV-Detected Radio Galaxy 3C~264: Core Emission and the Hot Accretion Flow Contribution}

\author[0000-0002-5267-2867]{Ka-Wah Wong} \affiliation{Department of Physics, SUNY Brockport, Brockport, NY 14420, USA}

\author{Colin M. Steiner}
\affiliation{Department of Physics, SUNY Brockport, Brockport, NY 14420, USA}

\author[0009-0000-9959-5216]{Allison M. Blum}
\affiliation{Department of Physics, SUNY Brockport, Brockport, NY 14420, USA}

\author[0000-0001-5683-5339]{Dacheng Lin}
\affiliation{Department of Physics, Northeastern University, Boston, MA 02115-5000, USA}

\author[0000-0003-3956-0331]{Rodrigo Nemmen}
\affiliation{Kavli Institute for Particle Astrophysics and Cosmology, Stanford University, Stanford, CA 94305, USA}
\affiliation{Instituto de Astronomia, Geof\'{\i}sica e Ci\^encias Atmosf\'ericas, Universidade de S\~ao Paulo, S\~ao Paulo, SP 05508-090, Brazil}

\author[0000-0003-4307-8521]{Jimmy A. Irwin}
\affiliation{Department of Physics and Astronomy, University of Alabama,
Box 870324, Tuscaloosa, AL 35487, USA}

\author[0000-0001-9110-2245]{Daniel R. Wik}
\affiliation{Department of Physics \& Astronomy, The University of Utah, 115 South 1400 East, Salt Lake City, UT 84112, USA}

\email{kwong@brockport.edu}


\begin{abstract}
3C~264 is one of the few FRI radio galaxies with detected TeV emission.  It is a low-luminosity AGN (LLAGN) and is generally associated with a radiatively inefficient accretion flow (RIAF).  Earlier multiwavelength studies suggest that the X-ray emission originates from a jet.  However, the possibility that the RIAF can significantly contribute to the X-rays cannot be ruled out.  In particular, hard X-ray emission $\gtrsim10$\,keV has never been detected, making it challenging to distinguish between X-ray models. Here we report a {\it NuSTAR} detection up to 25\,keV from 3C~264.  We also present subpixel deconvolved {\it Chandra} images to resolve jet emission down to $\sim$0\farcs2 from the center of the unresolved X-ray core.  Together with a simultaneous {\it Swift} observation, we have constrained the dominant hard X-ray emission to be from its unresolved X-ray core, presumably in its quiescent state. We found evidence of a cutoff in the energy around 20\,keV, indicating that at least some of the X-rays from the core can be attributed to the RIAF.  The Comptonization model suggests an electron temperature of about 15\,keV and an optical depth ranging between 4 and 7, following the universality of coronal properties of black hole accretion.  The cutoff energy or electron temperature of 3C~264 is the lowest among those of other LLAGNs.  The detected hard X-ray emission is at least an order of magnitude higher than that predicted by synchrotron self-Compton models introduced to explain $\gamma$-ray and TeV emission, suggesting that the synchrotron electrons might be accelerated to higher energies than previously thought.
\end{abstract}

\section{introduction}
\label{sec:intro}

Studying active galactic nuclei (AGNs) and their relativistic outflows or jets remains one of the most vibrant areas of research in astrophysics today.  It is critical not only for probing the properties of black holes but also for understanding their impact on the formation of larger-scale structures.  Some AGNs are identified as very high energy (VHE) TeV sources, indicating that particles within them are accelerated to equivalent energy scales.  Therefore, studying TeV-emitting AGNs provides crucial insights into the mechanisms of VHE radiation and acceleration \eg{Madejski2016}. 

Among the VHE sources discovered, blazars are the most common type of extragalactic sources, with their jets oriented closely to our line of sight, and are thought to be responsible for the TeV emission. 
A number of radio galaxies with misaligned jets have also been identified as TeV sources \citep{Rieger2018,Rani2019,Rulten2022}. 
In particular, four of these belong to the FRI type 
(NGC~1275, 3C~264, M87, and Cen A)
and seem to form a subclass of VHE sources\footnote{http://tevcat.uchicago.edu}.
Specifically, FRI radio galaxies typically host low-luminosity AGNs (LLAGNs; \citealt{Wu2007}), which are believed to accrete in a radiatively inefficient accretion flow (RIAF) or a hot mode \citep{Narayan1994, Yuan2014}.
Studying these TeV-emitting FRI sources will undoubtedly provide a unique scientific perspective on the study of VHE particle acceleration in AGNs, whether it be from a misaligned jet \citep{Abdo2010} or from the hot accretion flow.

3C~264 is a FRI radio galaxy located at a redshift of $z = 0.0217$ \citep{Baum1990}.  TeV emission was detected with VERITAS in 2018 \citep{Mukherjee2018}.
The mass of the supermassive black hole is about $2.6 \times 10^{8} M_{\odot}$
\citep{vandenBosch2016}.
The X-ray jet has been resolved with {\it Chandra}, alongside an unresolved X-ray core \citep[][hereafter, core means the central source unresolved by {\it Chandra} at the subarcsecond scale]{Perlman2010}.  The X-ray emission is dominated by the core with an X-ray luminosity of $L_{0.5-10\,{\rm keV}}=2\times10^{42}$ ergs s$^{-1}$ \citep{Sun2007}. Soft X-ray emission  below $\sim$2\,keV has been detected from the host galaxy NGC 3862, extending up to 6\arcsec, using {\it Chandra} \citep{Sun2007}. The coronal temperature has been measured to be $0.65^{+0.29}_{-0.09}$\,keV and the measured  coronal luminosity is $L_{0.5-2\,{\rm keV}}=1.4\times10^{40}$ ergs  s$^{-1}$.  The contribution of X-ray emission from LMXBs to the total X-ray emission is estimated to be at most 4\% \citep{Sun2002}. 

Due to its low Eddington ratio of $L_{\rm bol}/L_{\rm Edd}\approx7 \times 10^{-5}$, 3C 264 is a LLAGN \citep{Donato2004}, with the black hole accreting in the RIAF mode.
\citet{Donato2004} and \citet{Evans2006}
found significant correlations between the X-ray
core luminosity and the radio/optical luminosities, suggesting that
at least some of the X-rays come from a jet, although an origin from RIAF cannot be completely ruled out.
By measuring the X-ray, optical, and radio spectral slopes,
\citet{Perlman2010} argue that the X-ray emission is likely due
to the synchrotron process, which is consistent with a jet model.   
These observations suggest that
jets can be associated with RIAF as expected \citep{Yuan2014}.  
However, the contribution of X-rays from the RIAF itself remains a possibility in these observations. 
For instance, the photon index $\Gamma$ of the RIAF model, especially at very low accretion rates, can be similar to that of a jet model \citep{Nemmen2014}. 
Similarly, in another TeV FRI galaxy, Cen~A, the source of hard X-ray emission could be attributed to the RIAF, a jet, or a combination of both \citep{Furst2016}.

3C~264 is distinct from the other three VHE FRI galaxies in a number of
aspects.  For example, Cen~A and NGC~1275 have been detected with 
fluorescent Fe-K line,
indicating that X-ray emission can be dominated by accretion 
flows or coronae illuminating accretion disks
\citep[e.g.,][]{Fukazawa2015}, while none was detected from 3C~264 or
M87, suggesting that X-rays may be dominated by jets instead \citep[see, e.g.,][for the discussion of M87]{Wong2017}.  Additionally, M87 is
located in a hot gas rich environment (center of the Virgo Cluster),
whereas the host galaxy of 3C~264 is offset from its host galaxy cluster
\citep{Perlman2010}.
Thus, comparing 3C~264 with other VHE FRI galaxies
will provide  insights into the impact of the accretion 
flow/jet 
and the environment on the nature of the X-ray and TeV emissions.

3C~264 has been observed from radio to TeV energy \citep{Kagaya2017,Boccardi2019,Archer2020}.  The most favored model to explain the $\gamma$-ray and TeV emission is the jet scenario, with
synchrotron emission from a jet plus the inverse Compton emission
(synchrotron self-Compton, SSC).  The full spectral energy distribution (SED) from radio to $\gamma$-ray appears to be well explained by the SSC model.  However, prior to our study, hard X-ray emission above 10\,keV had not been detected.   In these SSC models, the hard X-ray region represents the transition between synchrotron and inverse Compton emissions. Therefore, modeling the full SED, including the hard X-ray spectrum, is crucial to identifying the emission origin.

In this paper, we report the detection of hard X-ray emission from 3C~264 above 10\,keV, using simultaneous {\it NuSTAR} and {\it Swift} observations. Additionally, we analyzed archival {\it Chandra} data to provide better constraints on the origin of the X-ray emission.  We describe the X-ray observations in Section~\ref{sec:obs}.  In Section~\ref{sec:image}, we present the image analysis of {\it NuSTAR} data, including deconvolved {\it Chandra} images, as well as an analysis of surface brightness and hardness profiles. Section~\ref{sec:spec} describes the spectral analysis results, focusing on the soft excess discovered with {\it Swift}, the overall hard X-ray emission from the AGN, and the decomposition the hard X-ray emission from the core and the jet.  We discuss the origin of the X-ray emission and compare our results with SSC models in Section~\ref{sec:discussion} and conclude in Section~\ref{sec:conclusion}.

We assume a flat universe with $H_0$\,=\,70\,km\,s$^{-1}$\,Mpc$^{-1}$,
$\Omega_M$\,=\,0.3, and $\Omega_{\Lambda}$\,=\,0.7. At the redshift of $z=0.0217$ \citep{Baum1990}, this gives 
an angular scale of 0.439\,kpc\,arcsec$^{-1}$ and a luminosity distance of 
94.5\,Mpc to 3C~264.  Errors are given at $1\sigma$ confidence level unless otherwise specified.

\section{X-ray Observations}
\label{sec:obs}

3C~264 was observed with {\it NuSTAR} on 2019 July 14, for 55ks (ObsID: 
60501016002). All data were reduced using the {\tt HEASoft} v6.31.1 and 
{\tt CALDB} version 20230530.
We reprocessed the data using the {\tt nupipeline} script of the 
NuSTAR Data Analysis Software package with standard screening 
criteria,
except that we set the filter mode using the parameter 
{\tt SAAMODE=optimized} in order to reduce the slightly 
elevated background caused by the South Atlantic Anomaly.

A {\it Swift} snapshot was taken during the {\it NuSTAR} observation on 
2019 July 15 for 1.9\,ks (ObsID: 00088885001). The {\it Swift-XRT} 
spectrum was extracted using online 
tools\footnote{\url{https://www.swift.ac.uk/user_objects/}} provided by 
the UK Swift Science Data Centre \citep{Evans2009}.

We also included a deep {\it Chandra} observation taken with the ACIS-S 
detector on 2004 January 24 (ObsID: 4916 for 38\,ks).
All the data were reprocessed using the {\it Chandra} Interactive Analysis 
of Observations ({\tt CIAO}) software version 4.13 and the {\it Chandra} 
{\tt CALDB} version 4.9.6.
The default subpixel event-repositioning algorithm ``{\tt EDSER}'' was 
used.
After removing the flares using the {\tt CIAO deflare} script, the 
final 
cleaned exposure time came to 34\,ks.

\begin{figure*}
\includegraphics[width=0.33\textwidth]{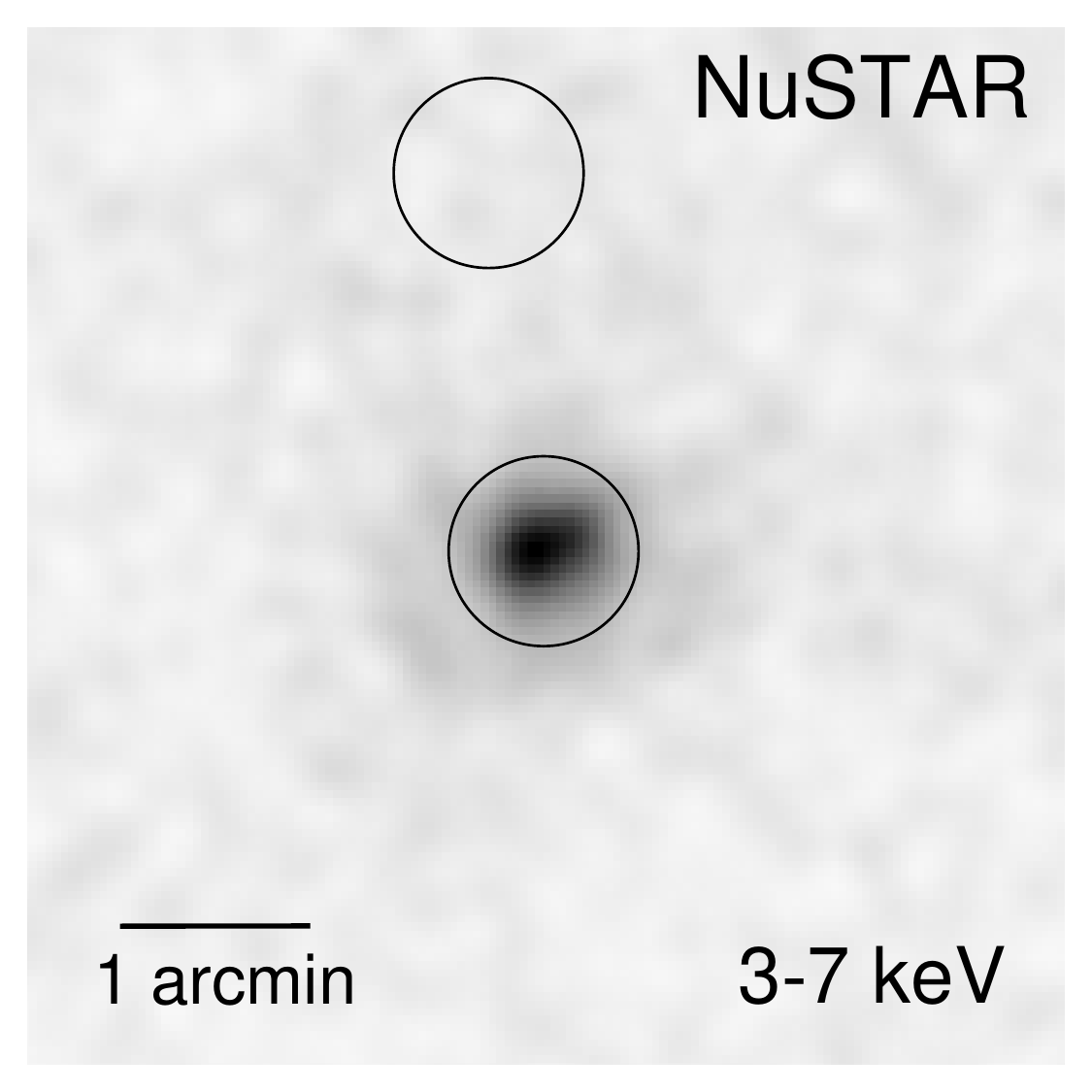}
\includegraphics[width=0.33\textwidth]{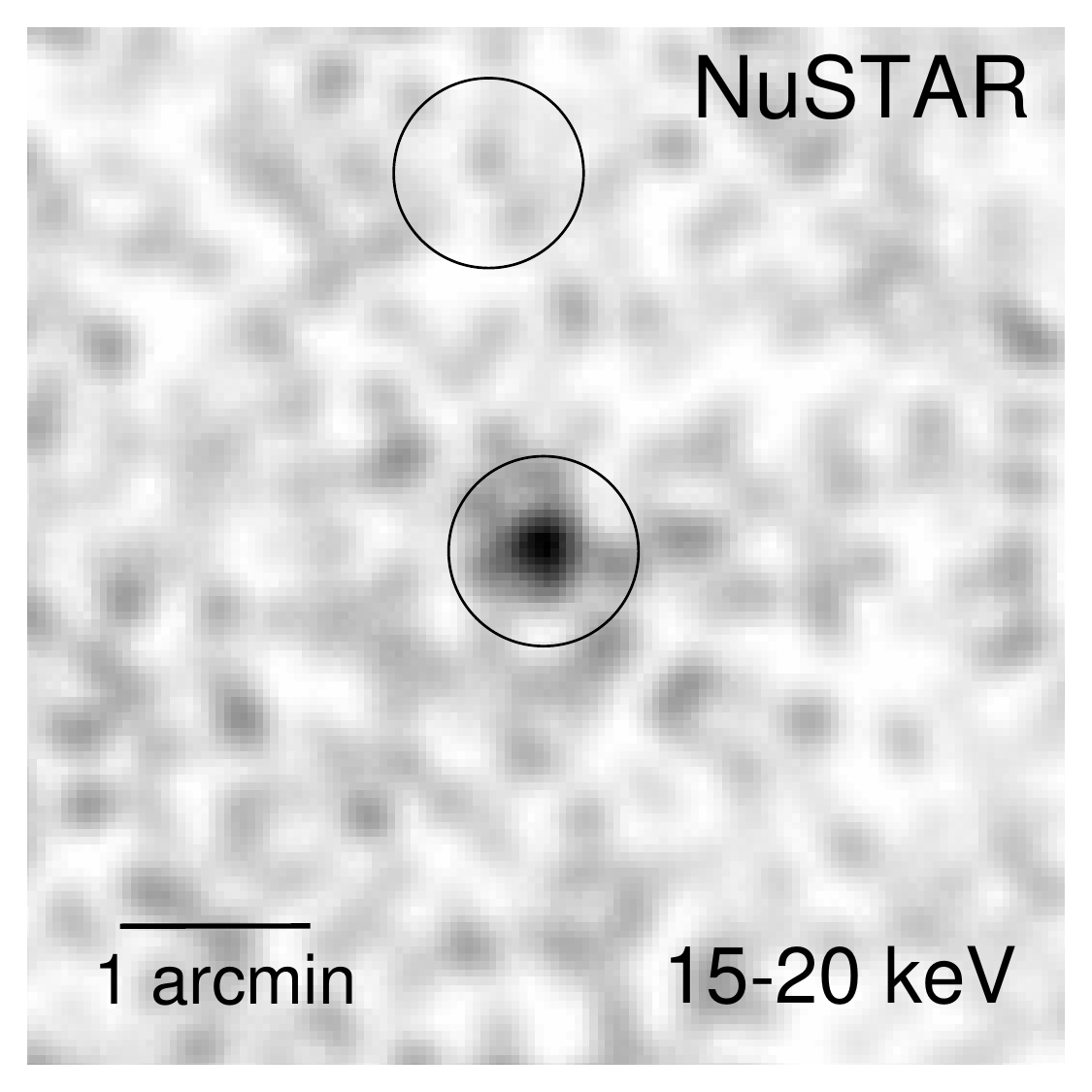}
\includegraphics[width=0.33\textwidth]{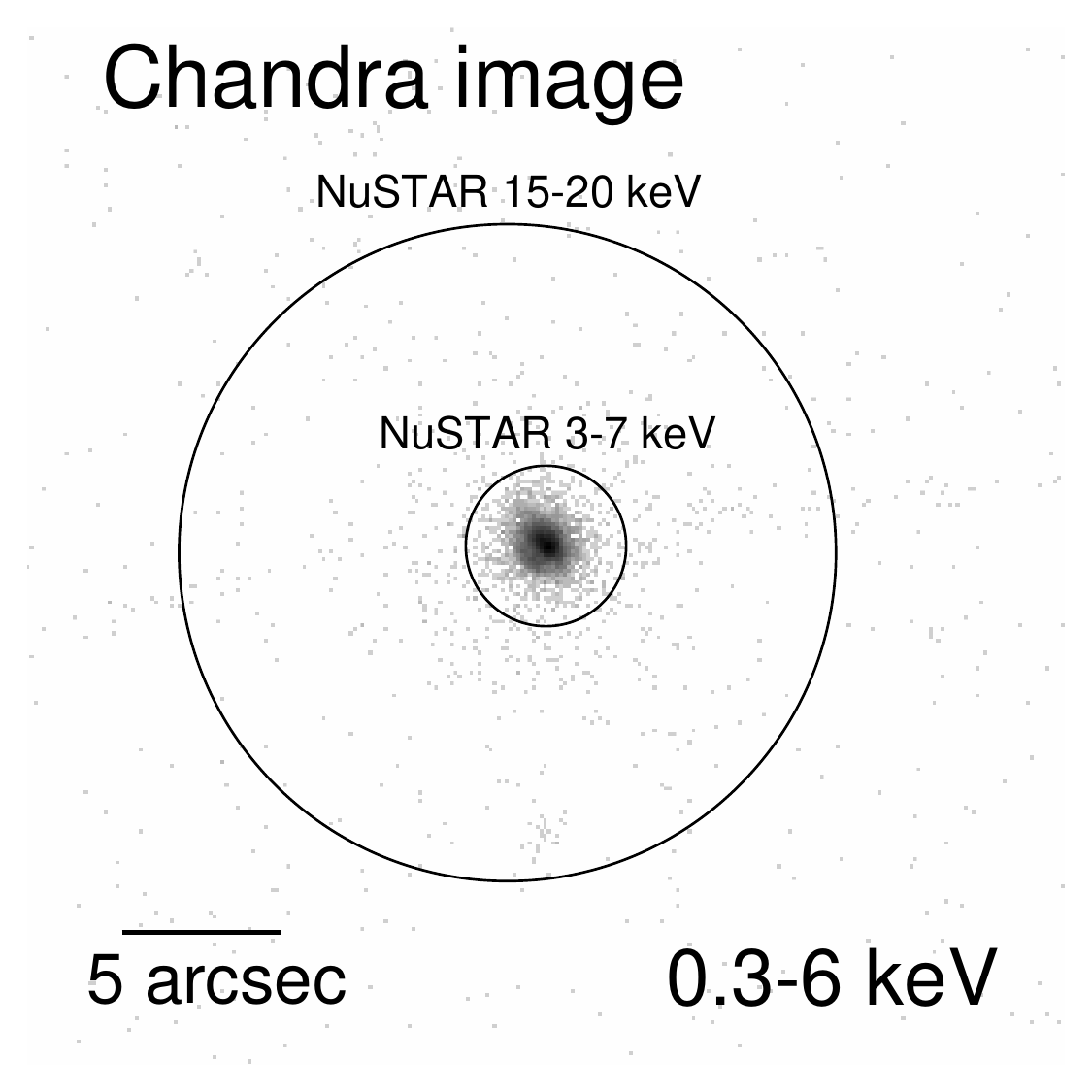}
\caption{Smoothed 3--7\,keV (left-hand panel) and 15--20\,keV (middle panel) {\it NuSTAR}
images of 3C~264.  The circle at the center of the image indicates the
30\arcsec\ radius spectral extraction region.  The circle on top is
the background spectral extraction region with the same radius of 
30\arcsec.  Using a larger background region gives essentially the same results.
A 0.3--6\,keV {\it Chandra} subpixel resolution image is shown in the 
right-hand panel, with one image pixel size = $0\farcs123$.
The two circles indicate the 3$\sigma$ position errors of
the emission peaks measured with {\it NuSTAR} in different energy bands.
Relative astrometric correction has been made so that the peak of the {\it 
NuSTAR} 3--7\,keV aligns with that of the {\it Chandra} emission (see the 
text).
The arc feature at about 0\farcs8 southeast from the center is an artifact 
of the {\it Chandra} PSF (Section~\ref{sec:image:decon}).
}
\label{fig:image}
\end{figure*}

\section{Image Analysis}
\label{sec:image}

\subsection{Spatial Analysis of {\it NuSTAR} Image}
\label{sec:image:nustar}

With the {\it NuSTAR} observation, hard X-rays above $\sim$10\,keV are 
clearly detected from 3C~264 for the first time. We explored images at 
different energy bands to determine the highest energy that could be 
detected.
Figure~\ref{fig:image} shows the {\it NuSTAR} images in 3--7\,keV and 
15--20\,keV.  Each image was created by combining the observations
from the two detectors.  Relative astrometry was corrected by matching 
the centroids of the 3--7\,keV emission.  Hard X-ray emission in 
15--20\,keV is clearly detected at 4$\sigma$.
In the higher energy band of 20--25\,keV, the detection significance of 
the source is at about 2$\sigma$.
No emission is detected at higher energy.

\begin{figure*}
\includegraphics[scale=0.5]{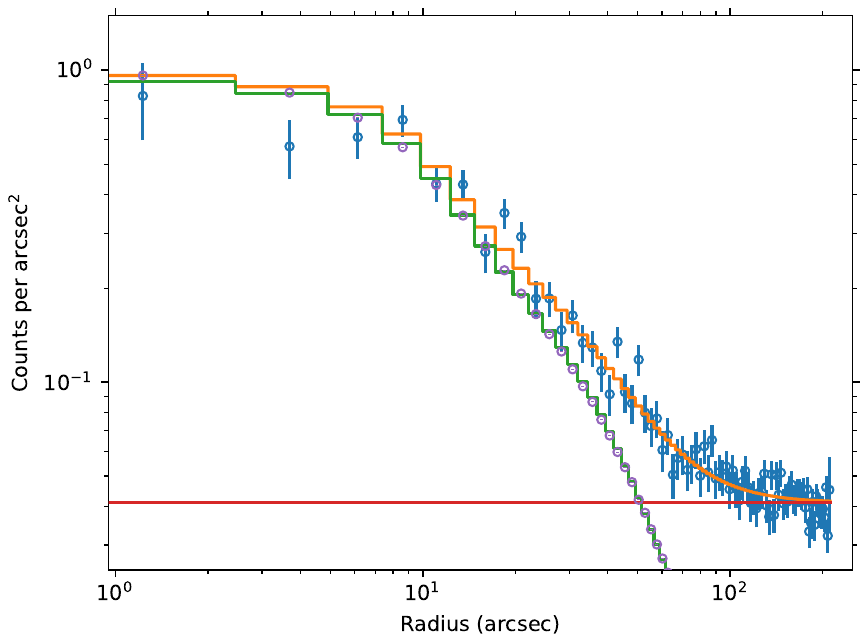}
\hspace{-3cm}
\includegraphics[scale=0.5]{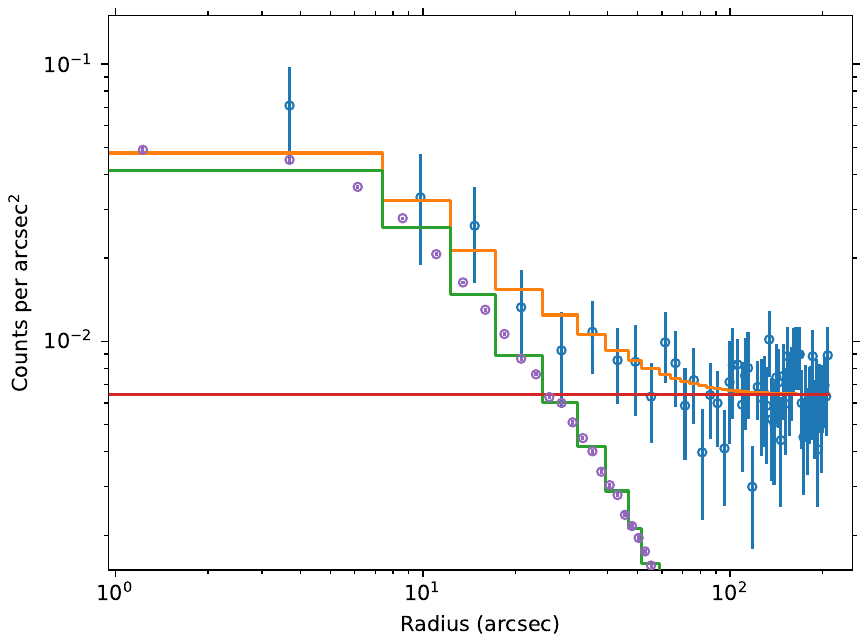}
\vspace{-1.5in}
\caption{Radial surface brightness profiles in the soft 
(3--7\,keV, left-hand panel) and hard (15--20\,keV, right-hand panel) bands.  {\it NuSTAR} data 
of 3C~264 are shown in blue circles.  
The Cyg~X-1 data, represented by purple circles with much smaller error 
bars, are normalized to match the surface brightness model of 3C~264.
The green and red lines represent the point source model and the 
background used to fit the 3C~264 data, respectively. 
The total emission profiles are shown in orange. The very good 
fitting indicates that both the soft and hard X-ray emissions are 
unresolved by {\it NuSTAR}.
}
\label{fig:profiles}
\end{figure*}

To quantify the spatial structure of the soft (3--7\,keV) and hard 
(15--20\,keV) X-ray
emission, and to test if the emissions are point-like or extended, we fit the 
images using {\tt CIAO}'s modeling and fitting package, {\tt Sherpa}. 
More specifically, we compared the spatial model with a known bright point 
source, Cyg~X-1, as observed with {\it NuSTAR} (ObsID: 10002003001). 
The data for Cyg~X-1 were processed in the same way as 3C~264 with standard screening criteria.
We fitted the NuSTAR soft and hard 
images of Cyg~X-1 with a 2D Gaussian model and a 2D $\beta$-model to 
represent 
the core and the wing of the PSF, respectively. 
The central coordinates of the models are tied.
A constant background 
model was also included in the fittings.
The fitting region is limited to a circular region of radius
$\sim$3\farcm5 centered near the peak of the emission.
This model fits the point source, Cyg~X-1, very well 
(Figure~\ref{fig:profiles}).
We use the best-fit 2D Gaussian model + 2D $\beta$-model as a point source 
template, along with a background model, to fit the images of 3C~264. The 
relative normalizations and all other parameters, except the 
central coordinates and 
the normalization of the 2D Gaussian model, are fixed.

Figure \ref{fig:profiles} shows that the point source model fits the 
images of 
3C~264 very well. We conclude that both the soft and hard X-ray emissions are 
unresolved by {\it NuSTAR}. 
The size of the 
X-ray jet resolved with {\it Chandra} is $\lesssim 2\arcsec$, 
which is much smaller than the 18\arcsec\ FWHM of the {\it NuSTAR} PSF.  
Thus, both the soft and hard X-ray emissions detected with {\it NuSTAR} are 
consistent with an origin that is similarly compact, presumably coming 
from the {\it Chandra} core and jet regions.

The locations of the {\it NuSTAR} soft and hard X-ray peaks can be 
determined with about 2\farcs5 and 10\farcs4 of precision at 3$\sigma$ for 
two parameters of interest, respectively.  The confidence regions are 
shown in the right-hand panel of Figure~\ref{fig:image}. Due to the 8\arcsec\ 
absolute astrometric uncertainty of {\it NuSTAR}, there is an offset of 
about 10\arcsec\ between the {\it Chandra} and {\it NuSTAR} soft peak.  
In the figure, a relative astrometric correction has been made so that the 
peak of the {\it NuSTAR} 3--7\,keV aligns with that of the {\it Chandra} 
emission. The locations of the {\it NuSTAR} soft and hard X-ray peaks are 
consistent with each other.  Thus, the origin of the hard X-ray emission 
is consistent with that of the soft emission (i.e., core/jet emission).

\subsection{{\it Chandra Image Deconvolution}}
\label{sec:image:decon}

Using a maximum entropy deconvolution algorithm, the X-ray jet has already 
been resolved with {\it Chandra} with the size of the jet of about 
2\arcsec\ pointing toward the northeast direct \citep{Perlman2010}.

To resolve the jet structure down to subarcsecond scale, we applied the 
Lucy-Richardson deconvolution algorithm to the {\it Chandra} subpixel 
resolution images \citep{Richardson1972,Lucy1974}.
For each energy band used, we first performed a raytracing simulation 
using the {\tt Chandra Ray Tracer 
(ChaRT)}\footnote{\url{https://cxc.harvard.edu/ciao/PSFs/chart2/index.html}}.  
The {\it Chandra} spectrum of 3C~264, fitted as an absorbed power-law 
model, and the aspect solution file from the {\it 
Chandra} observation were used for these raytracing simulations.
For each binning size of the image to be deconvoluted, a simulated event 
file was created by projecting the rays onto the 
detector plane using the {\tt CIAO simulate\_psf} script and the {\tt 
MARX} 
software\footnote{\url{https://space.mit.edu/cxc/marx/}}.
This script also created an image of the PSF.
Finally, the {\tt CIAO arestore} script was used, along with the PSF 
images, to perform the deconvolution.  

To test the convergence of the deconvolution, we experimented with various 
image binning sizes (0.0625, 0.1, 0.125, 0.25, and 1 physical pixel), as 
well as different numbers of iterations (50, 100, and 150). All of these 
trials yielded similar results, with the same notable features appearing 
in each image. As a result, we selected a binning size of 0.125 physical 
pixels and 100 iterations as our nominal choice.

\begin{figure*}
\includegraphics[width=0.5\textwidth]{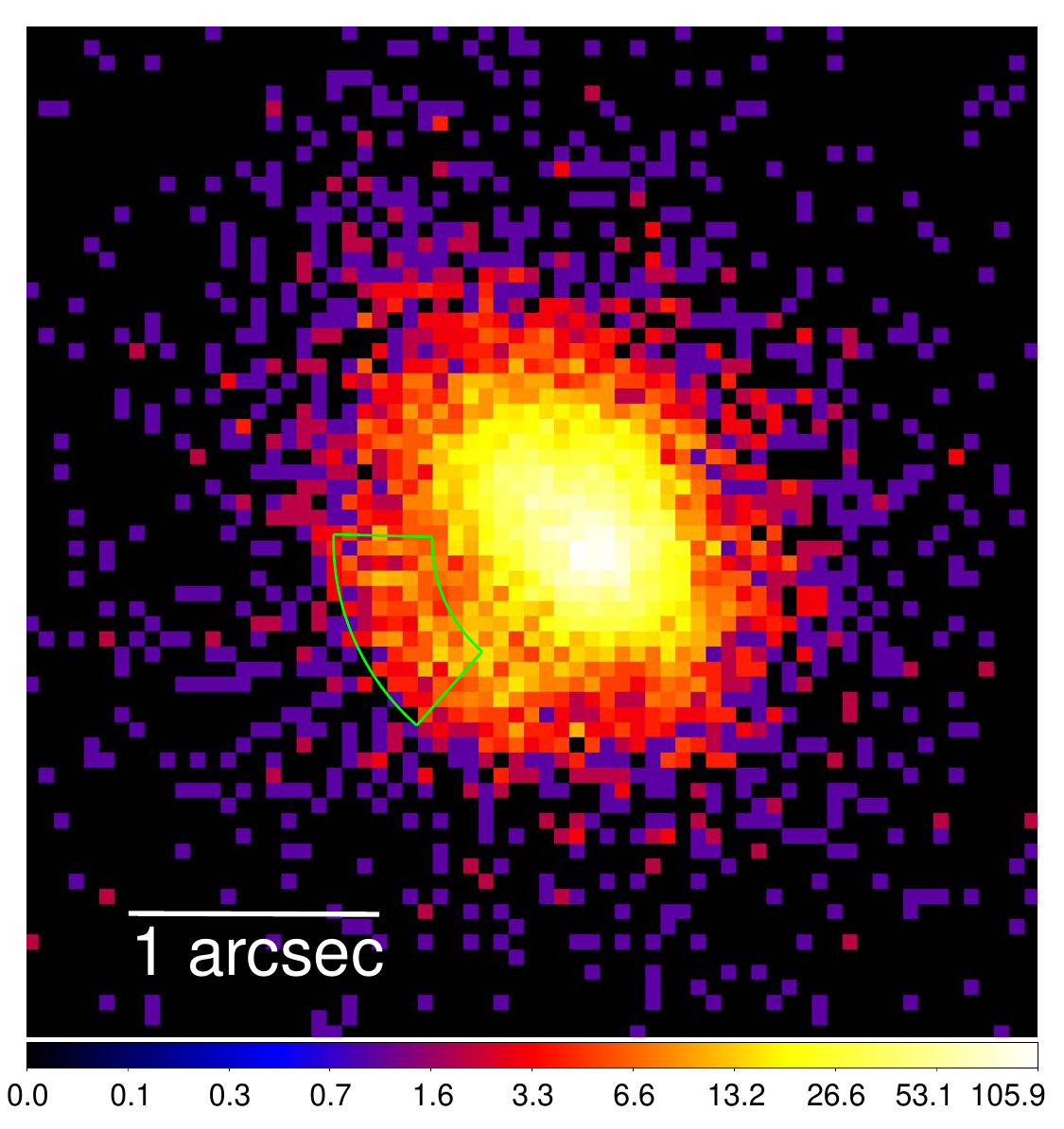}
\includegraphics[width=0.5\textwidth]{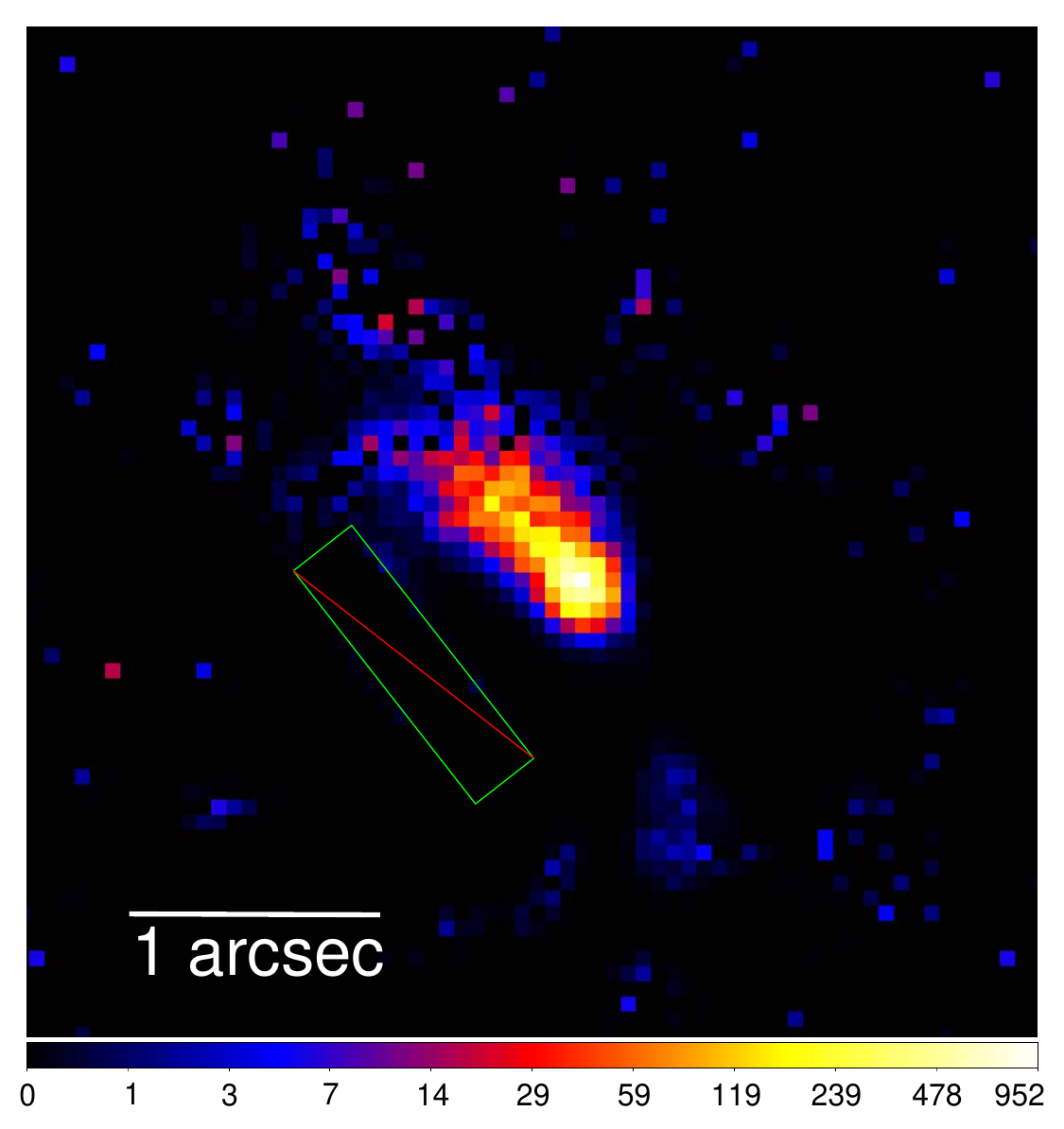}
\includegraphics[width=0.33\textwidth]{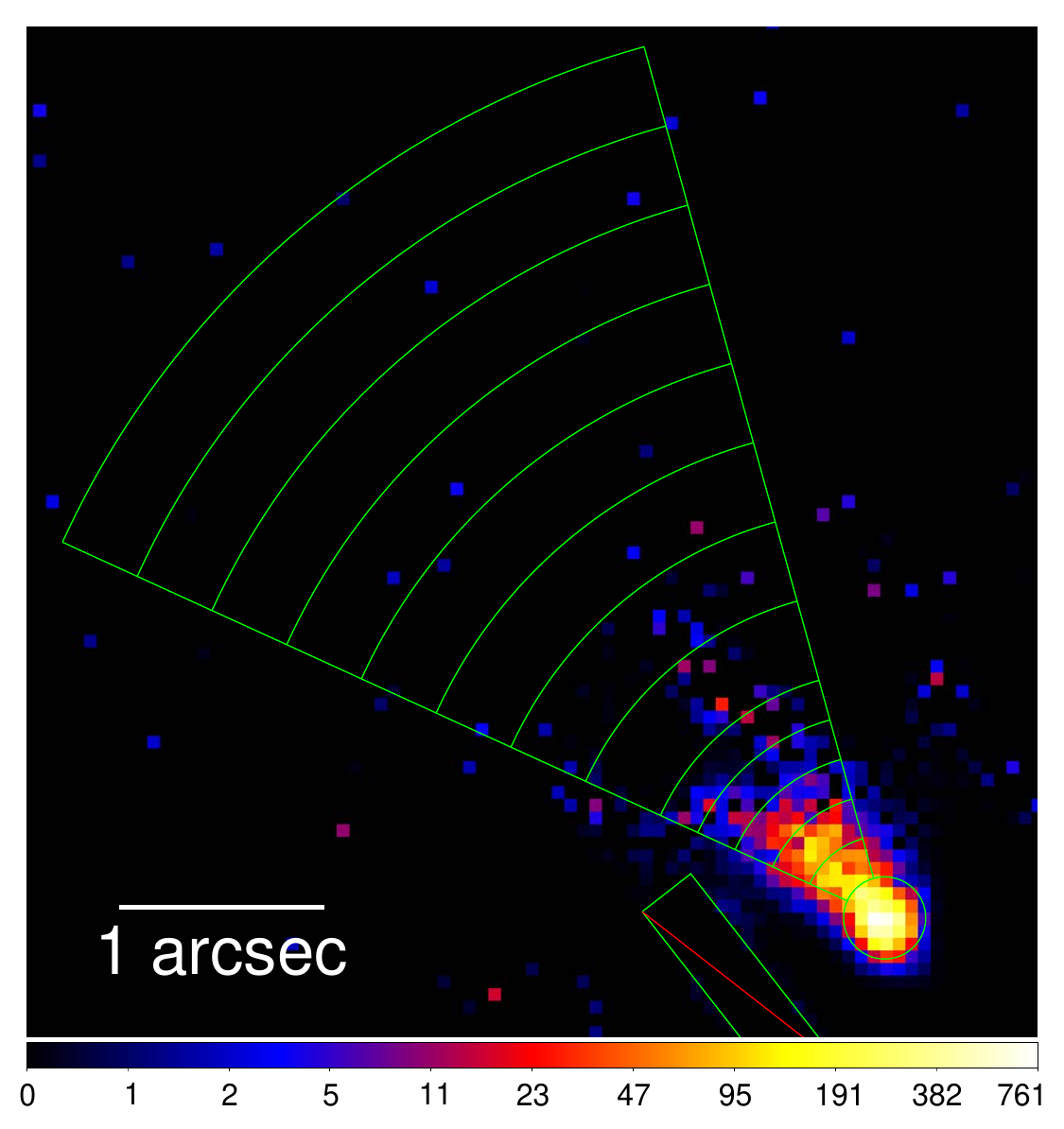}
\includegraphics[width=0.33\textwidth]{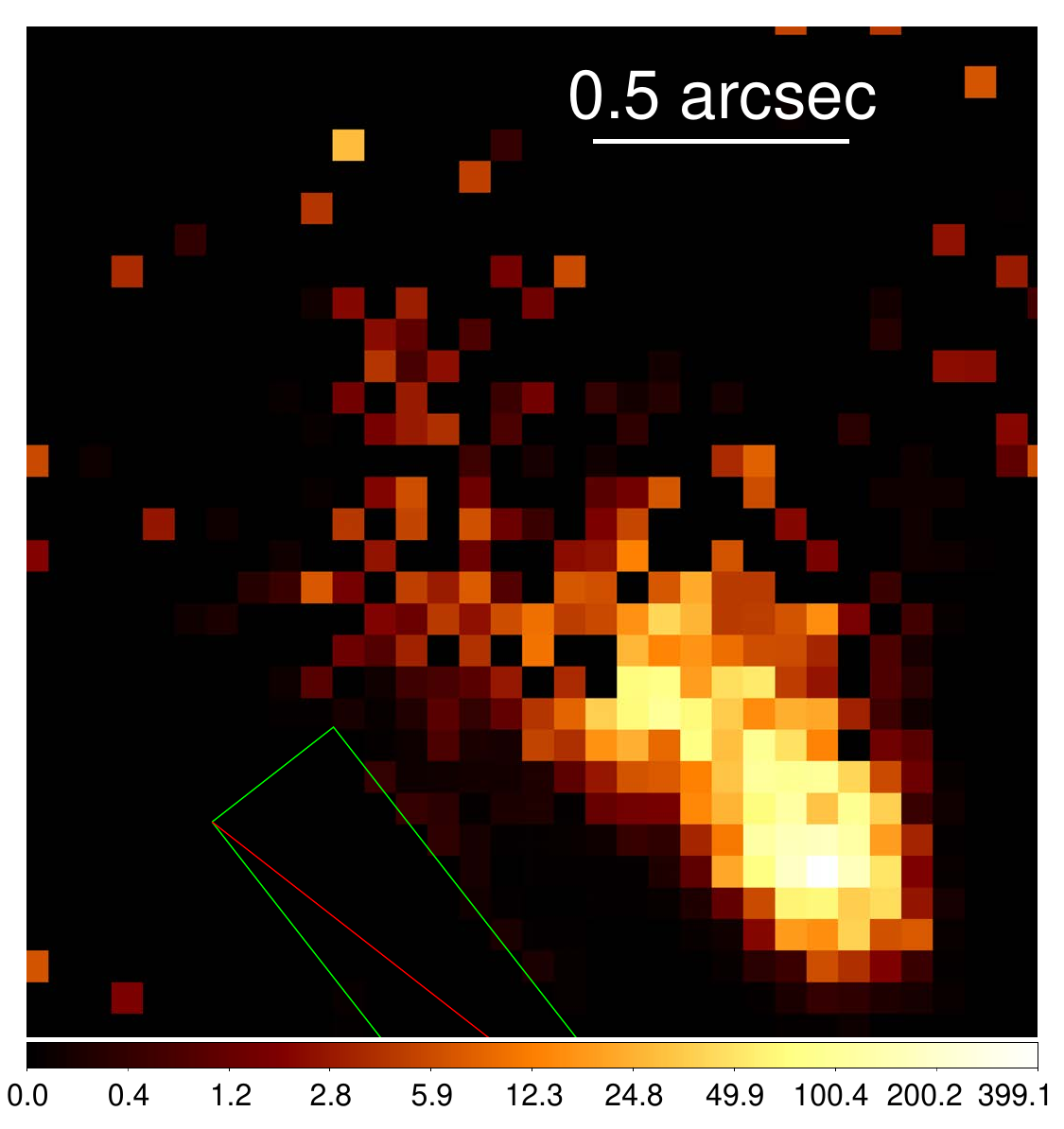}
\includegraphics[width=0.33\textwidth]{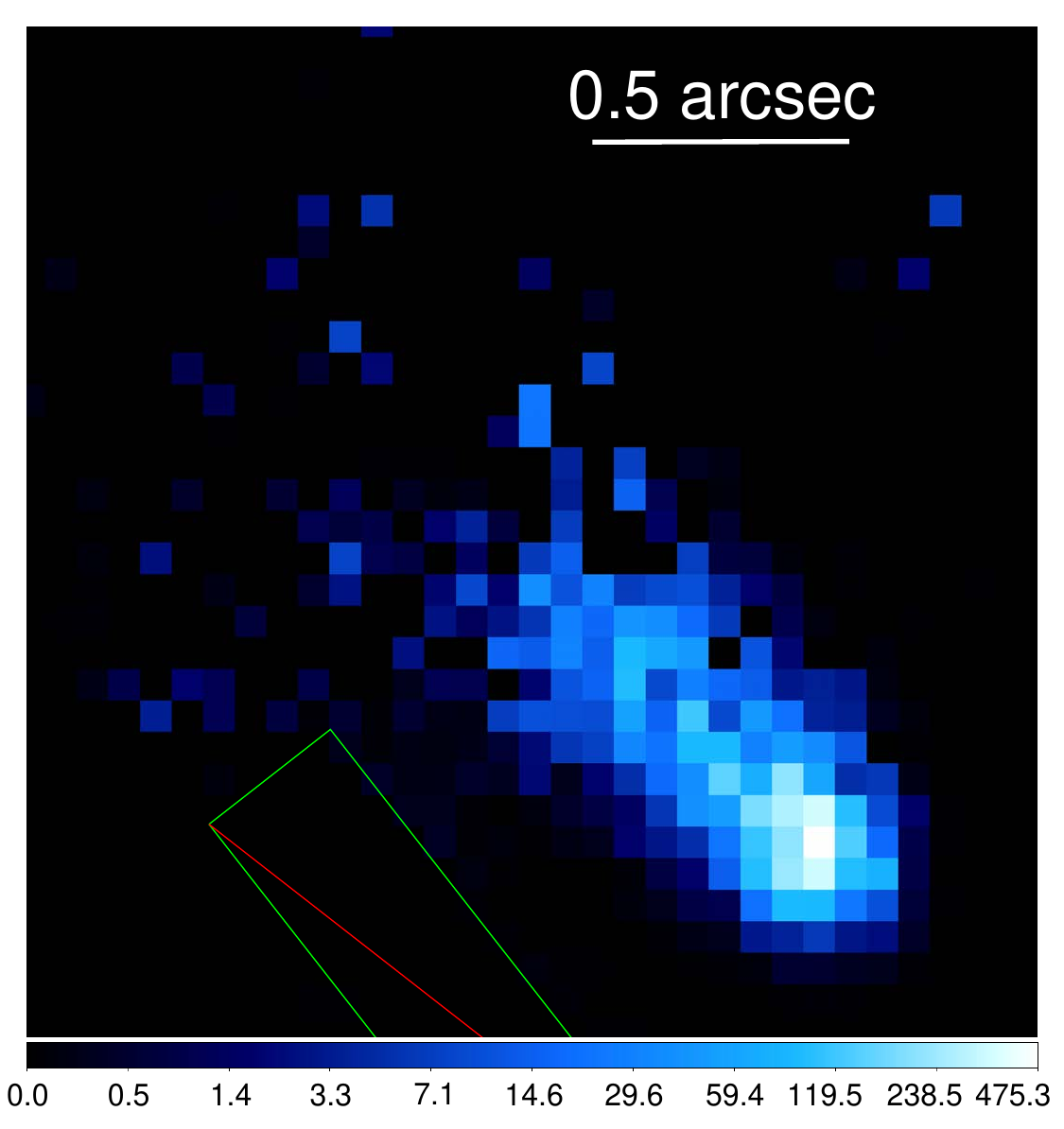}
\caption{Upper left-hand panel: 0.3--6\,keV {\it Chandra} subpixel resolution image.
The arc feature highlighted in the green 
region is an artifact of the {\it Chandra} PSF.  The emission is elongated 
in direction of the jet to the northeast.  
Upper right-hand panel: Deconvolved {\it Chandra} image in 0.3--6\,keV.  
Lower panels: Deconvolved {\it Chandra} image in the wide band 
(0.5--6\,keV; left-hand panel), soft band (0.5--1\,keV; middle panel), and hard band 
(1--6\,keV; right-hand panel).  The regions shown on the left 
are used for the analysis of surface brightness and hardness ratio 
profiles.
In all panels, the color bar units are counts per image pixel, and 
one image pixel size = 0\farcs0615.
For each deconvolved image, the PSF artifact, 
located within the box region, has been excluded.
}
\label{fig:image_arestore}
\end{figure*}

The upper panels in Figure~\ref{fig:image_arestore} show the high-resolution 
{\it Chandra} subpixel resolution image in 0.3--6\,keV and the 
corresponding deconvolved image.  The green region in the upper left-hand 
panel, created by the {\tt CIAO make\_psf\_asymmetry\_region} script, 
indicates the artifact of the 
PSF\footnote{\url{https://cxc.cfa.harvard.edu/ciao/caveats/psf_artifact.html}}. 
Aside from the artifact, the X-ray emission is elongated in the direction 
of the jet to the northeast.  Most of the emission is 
confined to an area of approximately 2\arcsec.  The deconvolved image 
reveals more details of the X-ray jet, showcasing a very narrow linear 
structure. The PSF artifact shows up in the same region and has been 
excluded on the image.

The jet is clearly one-sided with the core located to the southwest.  
Most of the X-ray emission originates from the core, which has a radius of 
$\sim$0\farcs2. The jet is most luminous within $\sim$0\farcs8, with the 
surface brightness decreasing progressively from the core. Beyond 
$\sim$0\farcs8, the emission has diminished significantly and the jet 
appears to fade away at about 2\arcsec.  The general spatial structure has 
been discussed by \citet{Perlman2010} using the maximum entropy 
deconvolution algorithm. However, the image deconvolved by the 
Lucy-Richardson method we used is noticeably sharper.  For instance, our 
deconvolved image shows two X-ray tails beyond $\sim$0\farcs8, one 
stronger to the northwest and the weaker one appears to turn west.  
Our deconvolved image is also sharper than that deconvolved with the {\it Chandra} HRC-I image observed in 2018 \citep{Archer2020}.

All these features are remarkably similar to the {\it HST} and {\it VLA} 
images, with much higher angular resolutions presented in 
\citet{Perlman2010}. This suggests that the deconvolution method is robust 
and that the jet emission originates from the same source across the radio 
to X-ray spectrum.

The lower three panels in Figure~\ref{fig:image_arestore} show the 
deconvolved images in three energy bands: the wide band in 0.5--6\,keV 
(lower left-hand panel), the soft band in 0.5--1\,keV (lower middle panel), and the hard band in 
1--6\,keV (lower right-hand panel).  
These figures were created for the analysis of surface brightness and 
hardness ratio  profiles that follows.
The general structures of the jets across the three bands are similar.  
However, the soft emission appears to be more extended, while the hard 
emission is narrower.

\subsection{Surface Brightness and Hardness Ratio Profiles of the Jet}
\label{sec:image:profiles}

\begin{figure*}
\includegraphics[width=0.5\textwidth]{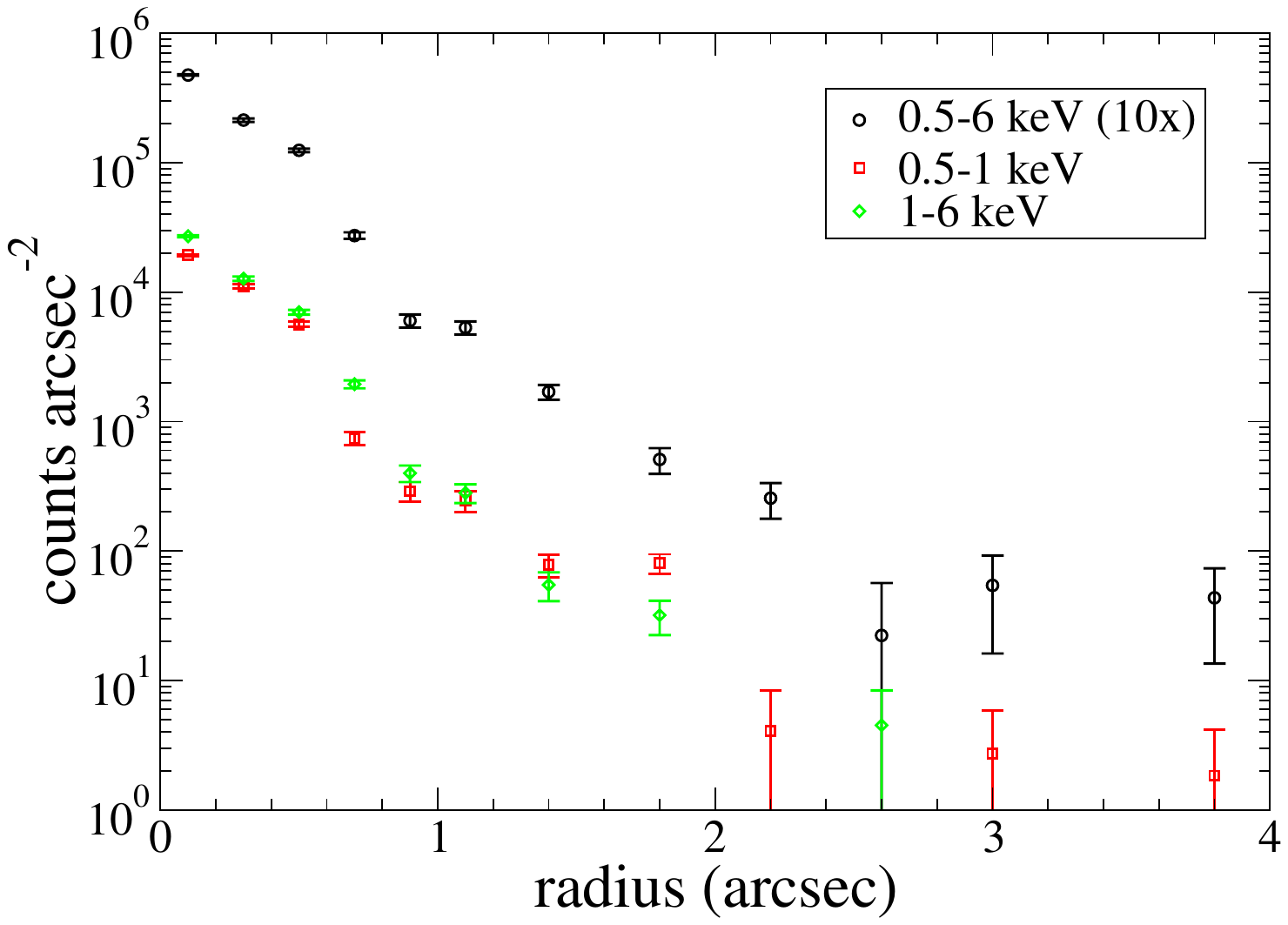}
\includegraphics[width=0.5\textwidth]{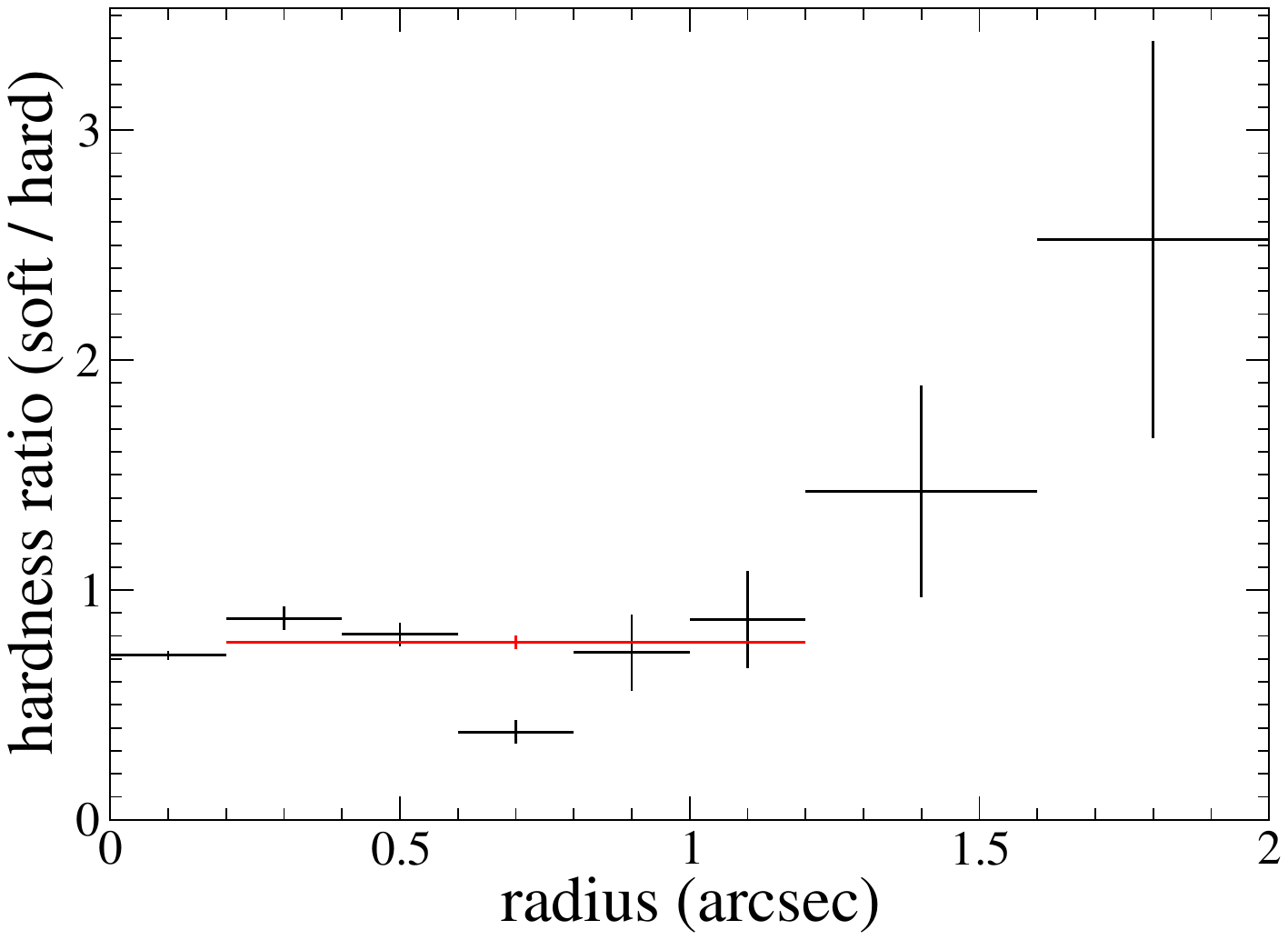}
\includegraphics[width=0.5\textwidth]{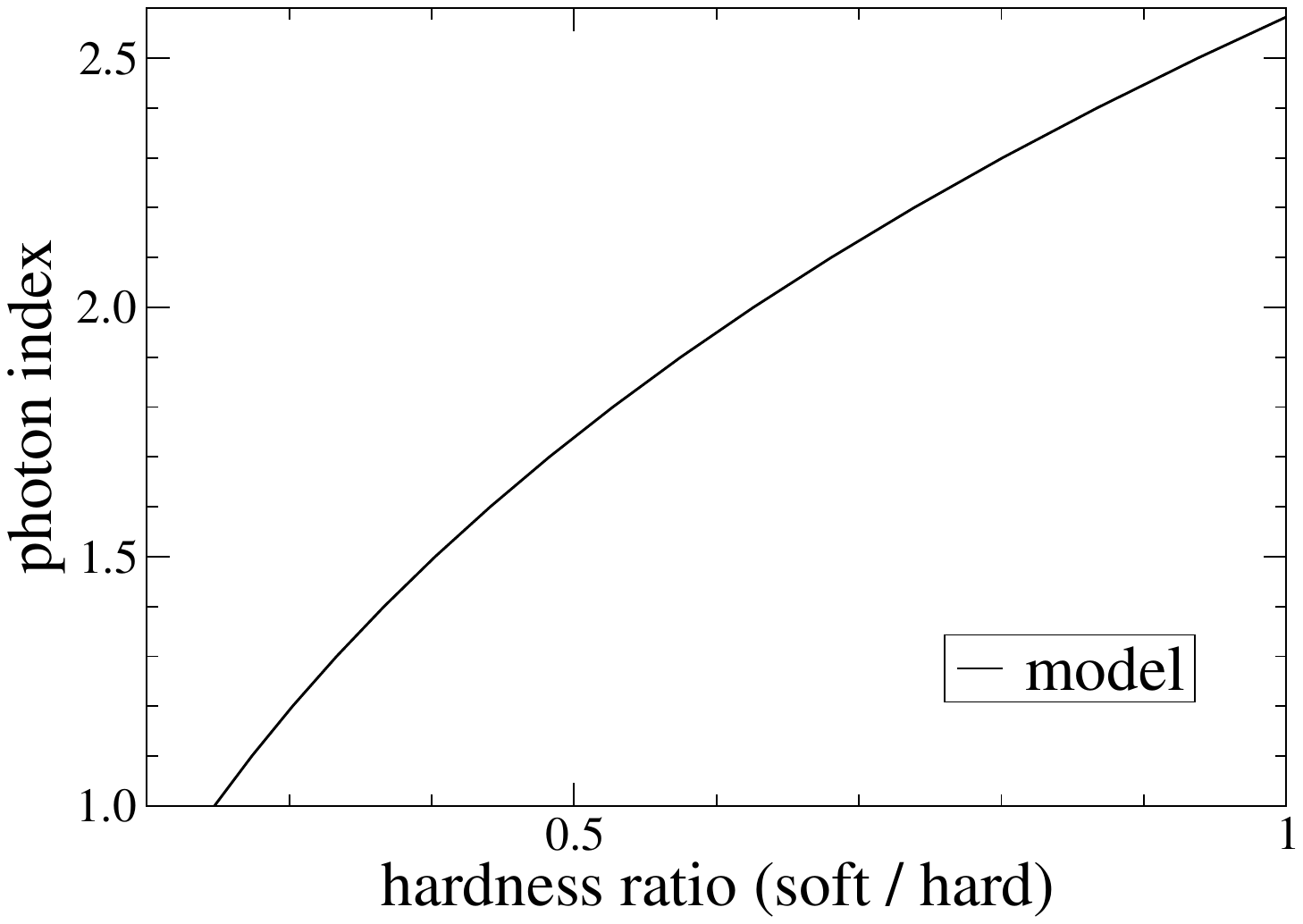}
\includegraphics[width=0.5\textwidth]{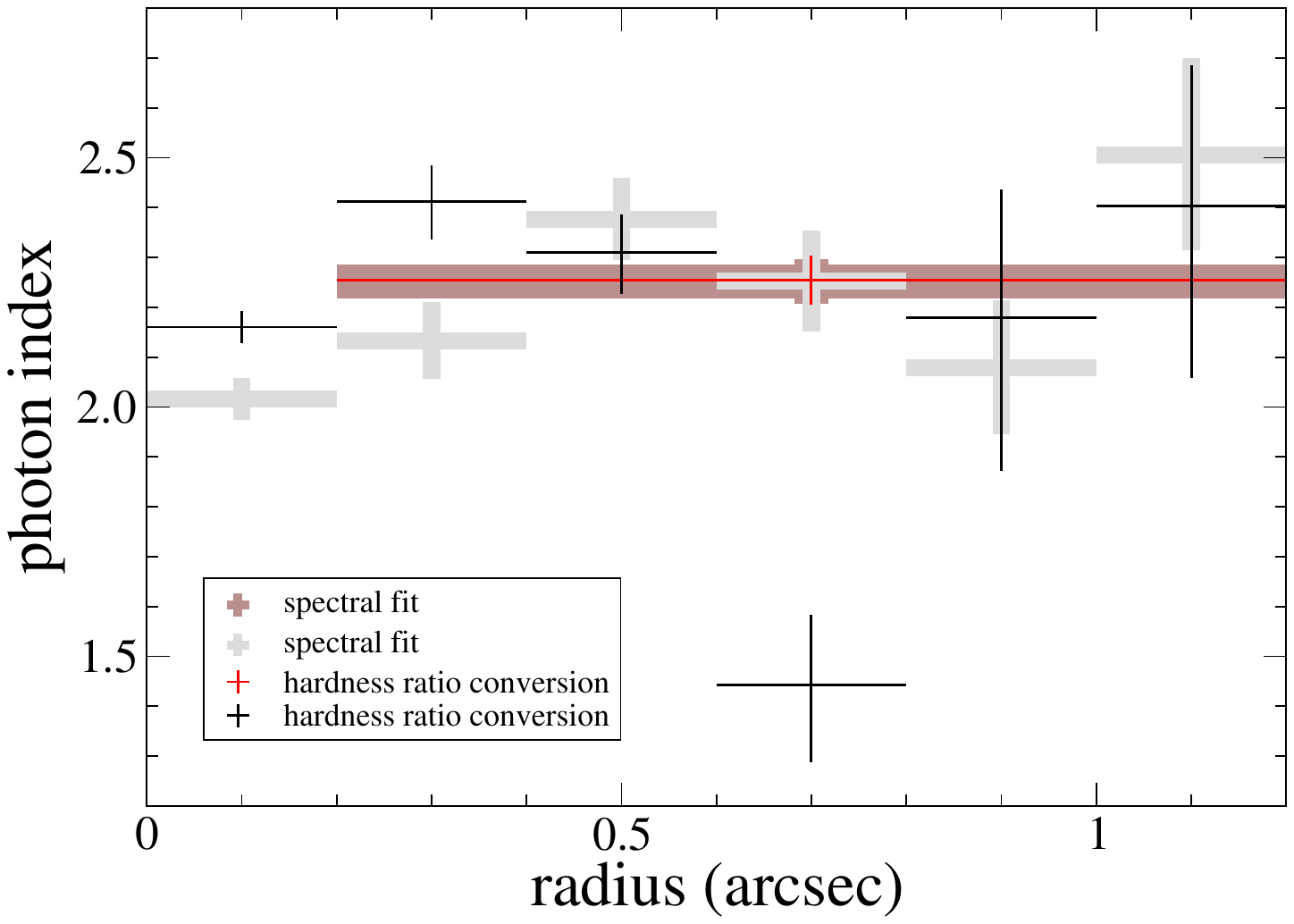}
\caption{Upper left-hand panel: 
Background-subtracted, exposure-corrected surface brightness profiles of
the jet in different energy bands.  The error bars are at 1$\sigma$ 
confidence level.  For clarity, the profile in 0.5--6\,keV is multiplied 
by a factor of 10.
Upper right-hand panel:
Hardness ratio profile defined as the ratio of the count in the soft 
energy band to the count in the hard energy band \citep{Park2006}.
The larger the hardness ratio, the softer the emission.
Errors are calculated using standard error propagation procedures.
The red data point, calculated using a large region, represents the 
average value across the brighter part of the jet.  The jet emission 
appears to be softer beyond 1\farcs2, but this could be due to soft 
emission from the galaxy or subject to systematic uncertainties.
Bottom left-hand panel:
Photon index vs. hardness ratio corresponding to the {\it Chandra} 
observation, assuming a power-law model absorbed by the Galactic column 
density.
Bottom right-hand panel:
Similar to the upper right-hand panel, but the hardness ratio has been 
converted to photon index (thin black and red crosses) using the figure in 
the bottom left-hand panel.  
The low photon index at 0\farcs7 appears to be an outlier (see text).
The photon indices determined using the spectral fitting 
method are also shown (thick gray and brown crosses).
}
\label{fig:profiles_chandra}
\end{figure*}

The upper left-hand panel in Figure~\ref{fig:profiles_chandra} displays the 
background-subtracted, exposure-corrected surface brightness profiles of 
the core and jet within 4\arcsec\ from the core center in three 
energy bands (wide, 
soft, and hard). The source regions used are depicted in the lower left-hand 
panel of Figure~\ref{fig:image_arestore}. An angular width of 50\arcdeg\ 
is chosen to encompass most of the X-ray emission observed within 
0\farcs8. The position angle of the central jet axis is set at 40\fdg43 
from north to east \citep{Meyer2015}. We selected a partial annulus 
background region with an angular width of 270\arcdeg\ extending from 
5\arcsec\ to 10\arcsec\ in the direction opposite the jet. Vignetting 
correction has been applied using the corresponding exposure maps.  
The energy bands have been chosen to optimize the hardness ratio analysis 
that follows.
Errors are estimated using the Gehrels prescription\footnote{Uncertainty 
$\sigma = \sqrt{N+0.75}+1$, where $N$ is the number of counts.} 
\citep{Gehrels1986}.

Beyond the unresolved core, approximately 0\farcs2 in radius, the surface 
brightness rapidly decreases out to about 0\farcs8, where the most obvious 
part of the jet lies. Past this point, the jet becomes quite faint and the 
surface brightness decreases more gradually up to about 2\farcs4. 
Any further out, the emission is consistent with background.

The deconvolved images from the soft and hard bands enable us to 
investigate the rough spectral properties at a spatial resolution higher 
than that achievable through spectral analysis. The latter is limited by 
PSF blurring. To accomplish this, we performed a hardness ratio analysis, 
a technique that has been widely used to characterize various X-ray 
astrophysical sources 
\citep[see, e.g.,][]{Sarazin2001,Park2006,Wong2008,Thimmappa2022}.

We experimented to determine the optimal energy bands for the hardness 
ratio analysis, and two energy bands were chosen: the soft (0.5--1\,keV) 
and hard (1--6\,keV) bands.  The lower limit of 0.5\,keV was chosen to 
minimize 
uncertainty in the Galactic column density, which more significantly 
impacts soft photons.  
Additionally, when fitting the {\it Chandra} spectrum within the central 
2\arcsec\ region, we observed residuals around 0.3\,keV. This makes the 
interpretation of energies lower than $\sim$0.5\,keV challenging.
The upper limit at 6\,keV was set to maximize the signal-to-noise ratio. 
The energy value of 1\,keV, which serves as the midpoint between our 
selected energy bands, is chosen to ensure that the signal-to-noise ratio 
in each region is greater than three within a 2\arcsec\ radius.

The upper right-hand panel in Figure~\ref{fig:profiles_chandra} displays the 
hardness ratio profile, which is derived by dividing the net photon counts 
in the soft band by those in the hard band for each radial region, or 
equivalently dividing the soft surface brightness profile by the hard 
surface brightness profile. The larger the hardness ratio, the softer the 
emission. The general trend suggests that the hardness ratio remains flat 
within 
$\sim$1\farcs2 
and softens from there to $\sim$2\arcsec, with an 
anomalous data point at 0\farcs7 displaying particularly hard emission. 
This anomalous data point is due to the abrupt decrease in soft photons in that region (see the upper left-hand panel in Figure~\ref{fig:profiles_chandra}).  The surface brightness profiles change significantly around that radius, which could introduce a large systematic uncertainty in that region.
The hardness ratio beyond 
$\sim$1\farcs2 
seems excessively soft for jet 
emission, which could be 
due to soft emission from the galaxy or systematic uncertainties. More 
data is needed to confirm whether this rising trend and the anomalous 
point are real. 

Upon closer inspection of the hardness ratio profile, the unresolved core 
within 0\farcs2 is significantly harder than the jet region beyond that at 
0\farcs2--0\farcs4. 
This does not seem to be due to the systematic error 
related to the PSF's dependence on energy because a larger PSF at higher 
energy should only make the outer regions harder, not the core. Spectral 
fitting also shows that the central core is harder than the jet on 
average, although the spatial resolution of the spectra is limited by the 
PSF. In principle pileup can make the central spectrum harder, but we 
have checked this with a {\tt PILEUP} model in {\tt XSPEC}\footnote{The X-ray Spectral Fitting Package
({\tt XSPEC}), \url{https://heasarc.gsfc.nasa.gov/xanadu/xspec/}} and found that 
it can bias the photon index ($\Gamma$) by at most 0.6\%, which is 
negligible.

To convert the hardness ratio into the photon index of the 
absorbed power-law model, we calculate the ratio of model-predicted rates 
in the soft and hard bands with different photon indices using {\tt 
XSPEC}.  Galactic absorption was assumed and the corresponding response 
files were used.  The relationship between the photon index and the 
hardness ratio is shown in the lower left-hand panel of 
Figure~\ref{fig:profiles_chandra}. The converted photon index profile 
within the brighter part of the jet ($<1\farcs2$) is shown as black 
crosses in the lower 
right-hand panel of Figure~\ref{fig:profiles_chandra}. 
The photon index at the 
$<0\farcs2$ core is $\Gamma=2.16^{+0.03}_{-0.03}$, which is significantly 
harder than that in the jet region of 0\farcs2--0\farcs4, where 
$\Gamma=2.41^{+0.07}_{-0.08}$. We also measured the average hardness ratio 
of the jet between 0\farcs2 and 1\farcs2, and found that the average photon 
index of the jet is $\Gamma=2.26^{+0.05}_{-0.05}$ (red cross in the 
right-hand panels of Figure~\ref{fig:profiles_chandra}).
Including the core from 0\arcsec to 1\farcs2, the average photon 
index is $\Gamma=2.19^{+0.03}_{-0.03}$ (not shown on the figure for clarity).

We also created spectra in the corresponding surface brightness regions 
and performed fitting to obtain the best-fit photon indices (thick crosses 
in the lower right-hand panel of Figure~\ref{fig:profiles_chandra}). Since the 
region sizes are smaller than the PSF, the spectra across regions are 
mixed. Nevertheless, the general trend of the photon indices obtained by 
spectral fitting and the magnitude of the uncertainties are consistent 
with those obtained by the hardness ratio method. 
In particular, the average spectral index of the jet between 0\farcs2 and 
1\farcs2 determined by the two methods are almost the same, suggesting 
that the hardness ratio analysis is reliable.

Note that the innermost two regions are harder according to the spectral 
fitting, while only the innermost region appears harder in the hardness 
ratio analysis. This discrepancy might be due to spectral mixing in the 
spectral analysis, whereas for the hardness ratio analysis deconvolved 
images are used. The spectral analysis does not confirm the particularly 
hard region at 0\farcs7, which could possibly be attributed to spectral 
mixing.

\section{Spectral Analysis}
\label{sec:spec}

\begin{figure*}
\centering
\includegraphics[width=0.45\textwidth, clip, trim={0.3cm 0.2cm 2.8cm 2.8cm}]{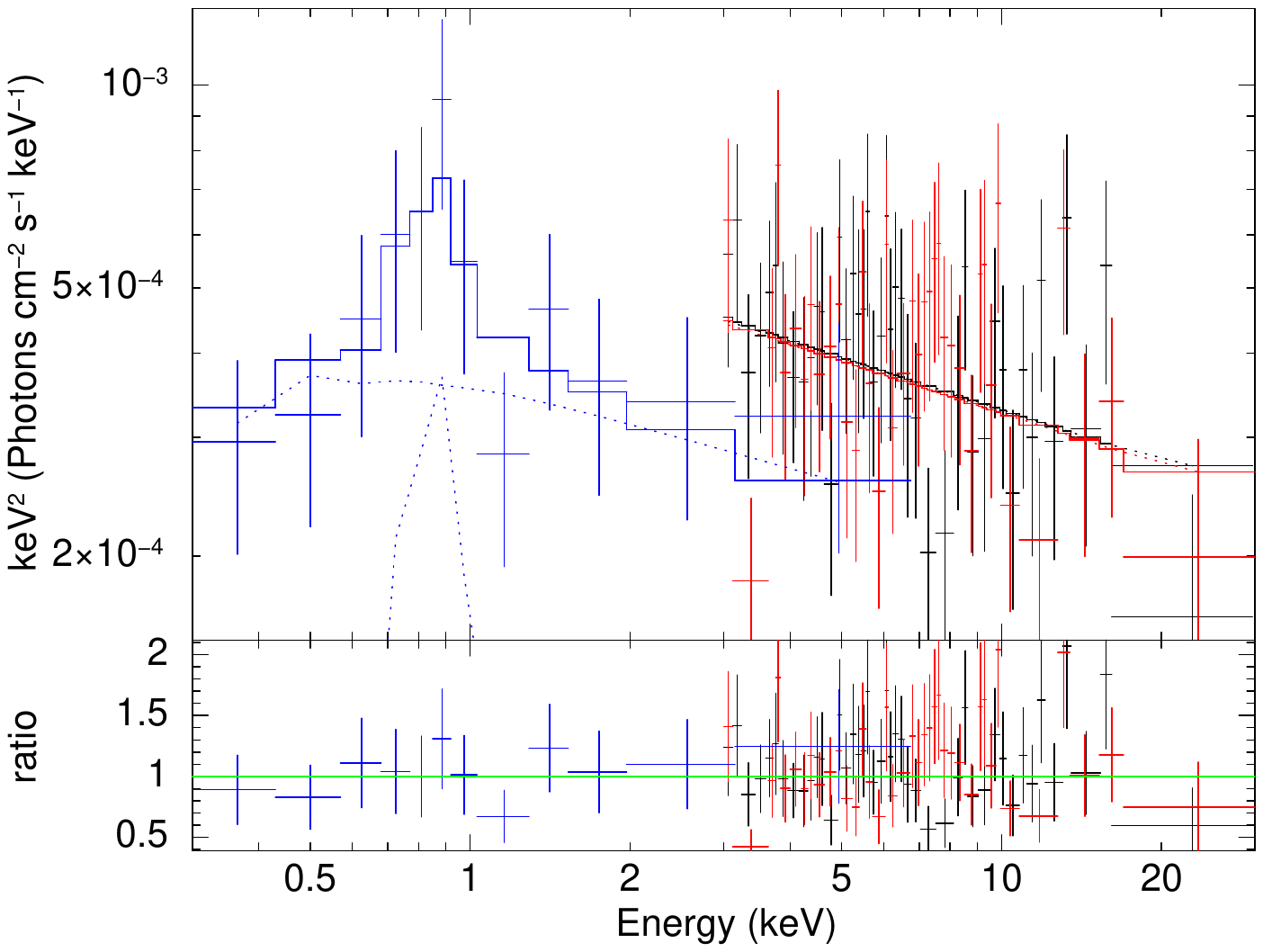}
\includegraphics[width=0.45\textwidth, clip, trim={0.3cm 0.2cm 2.8cm 2.8cm}]{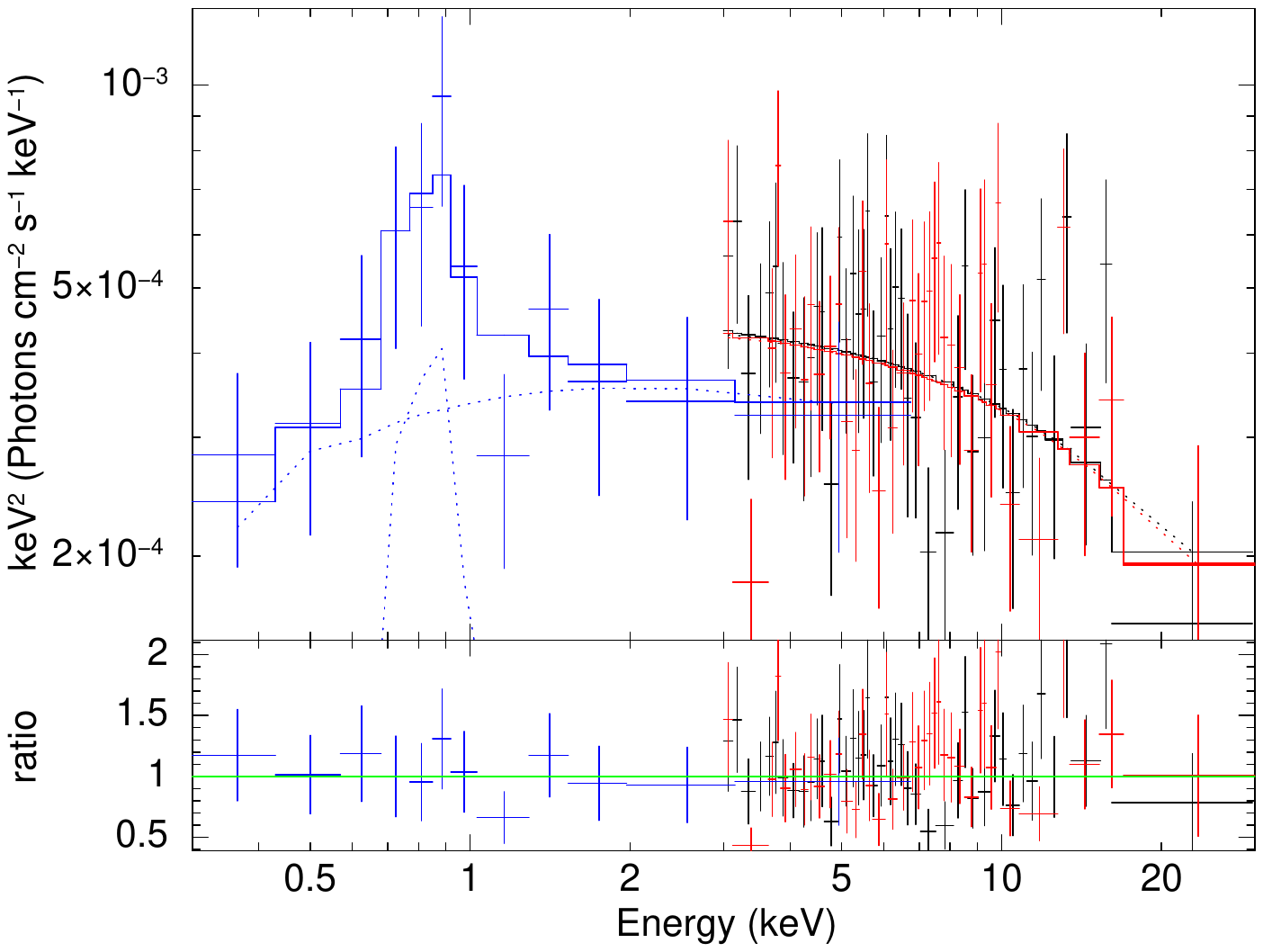}
\includegraphics[width=0.45\textwidth, clip, trim={0.3cm 0.2cm 2.8cm 2.8cm}]{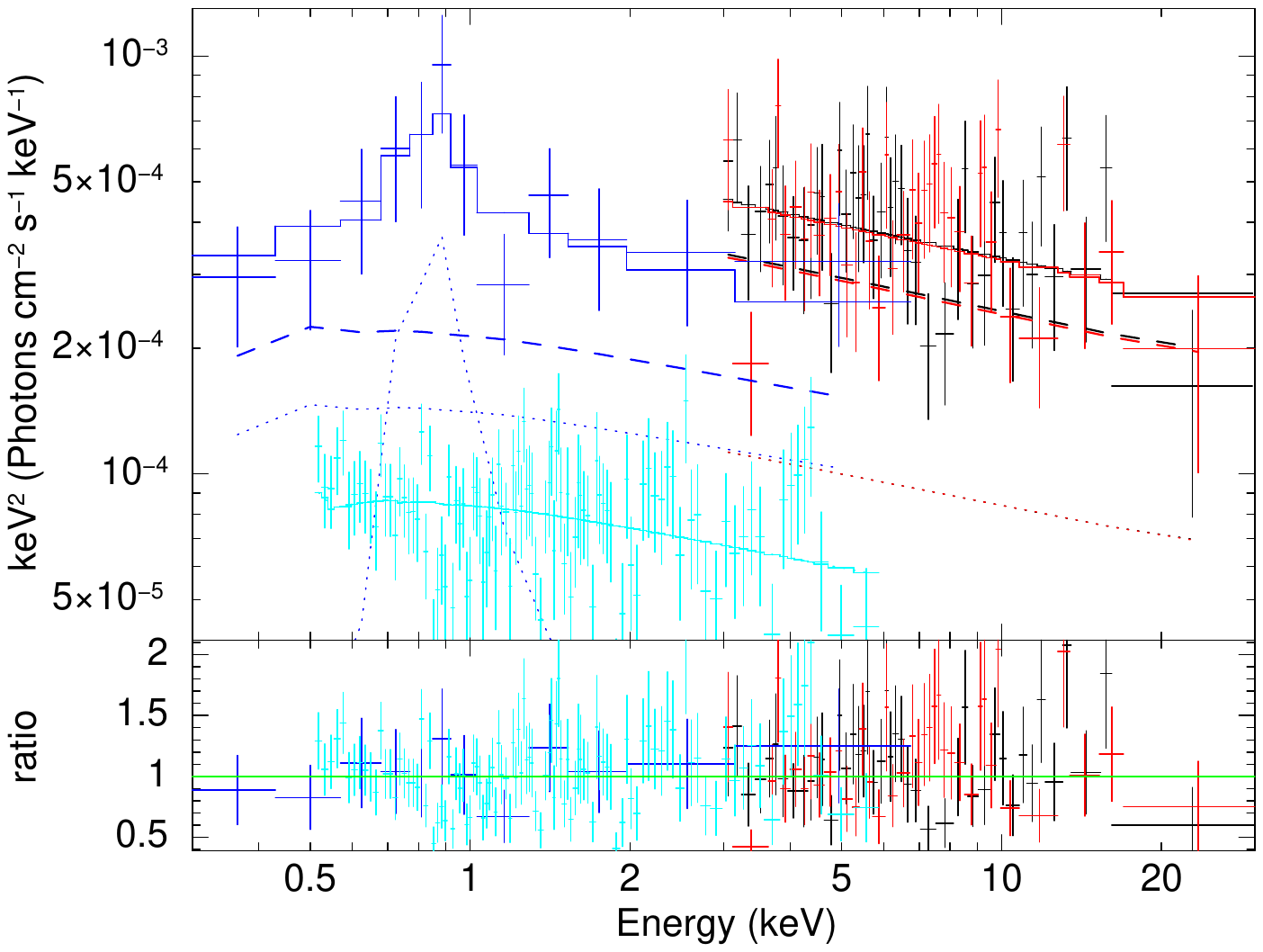}
\includegraphics[width=0.45\textwidth, clip, trim={0.3cm 0.2cm 2.8cm 2.8cm}]{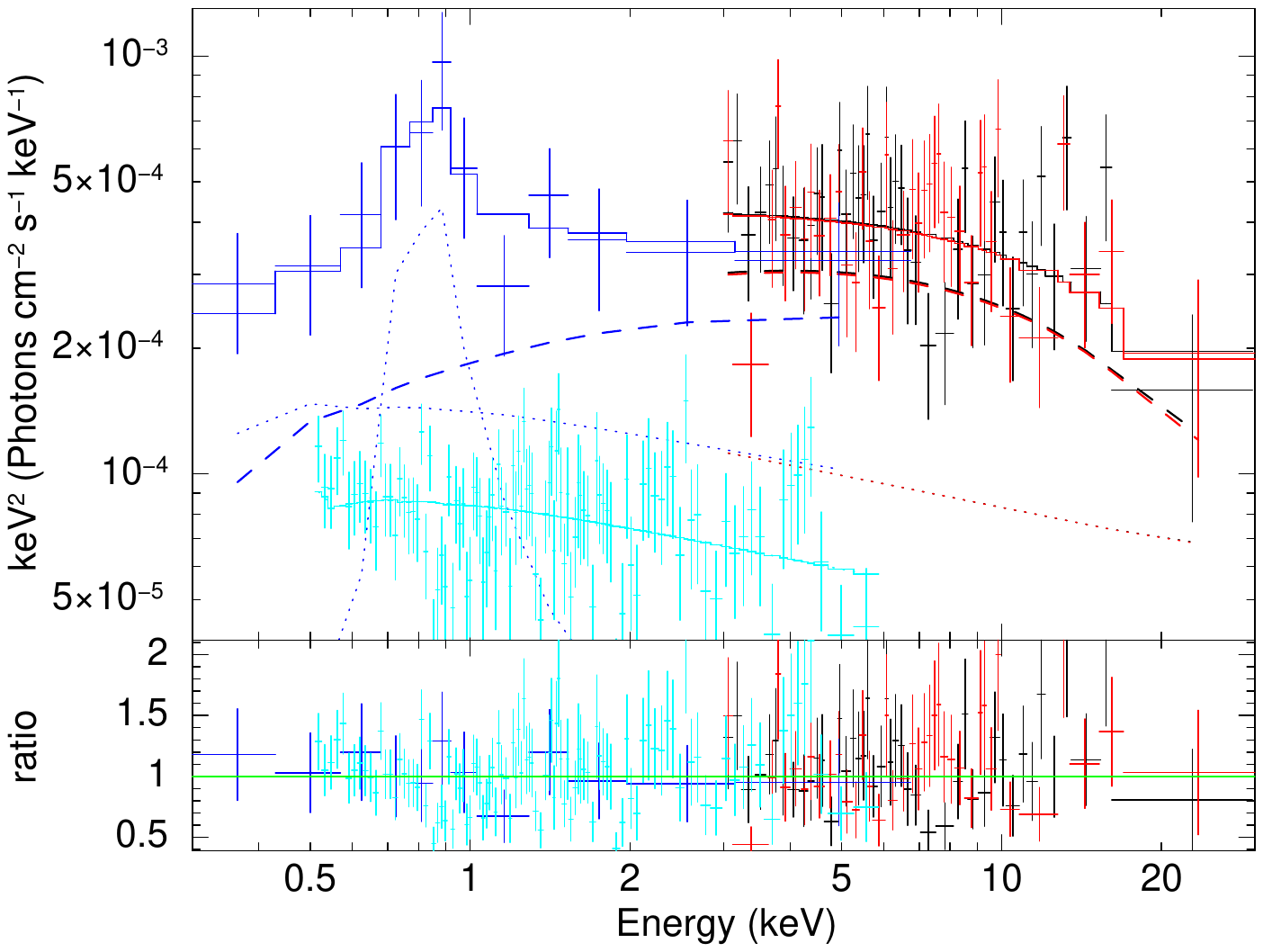}
\includegraphics[width=0.45\textwidth, clip, trim={0.3cm 0.2cm 2.8cm 2.8cm}]{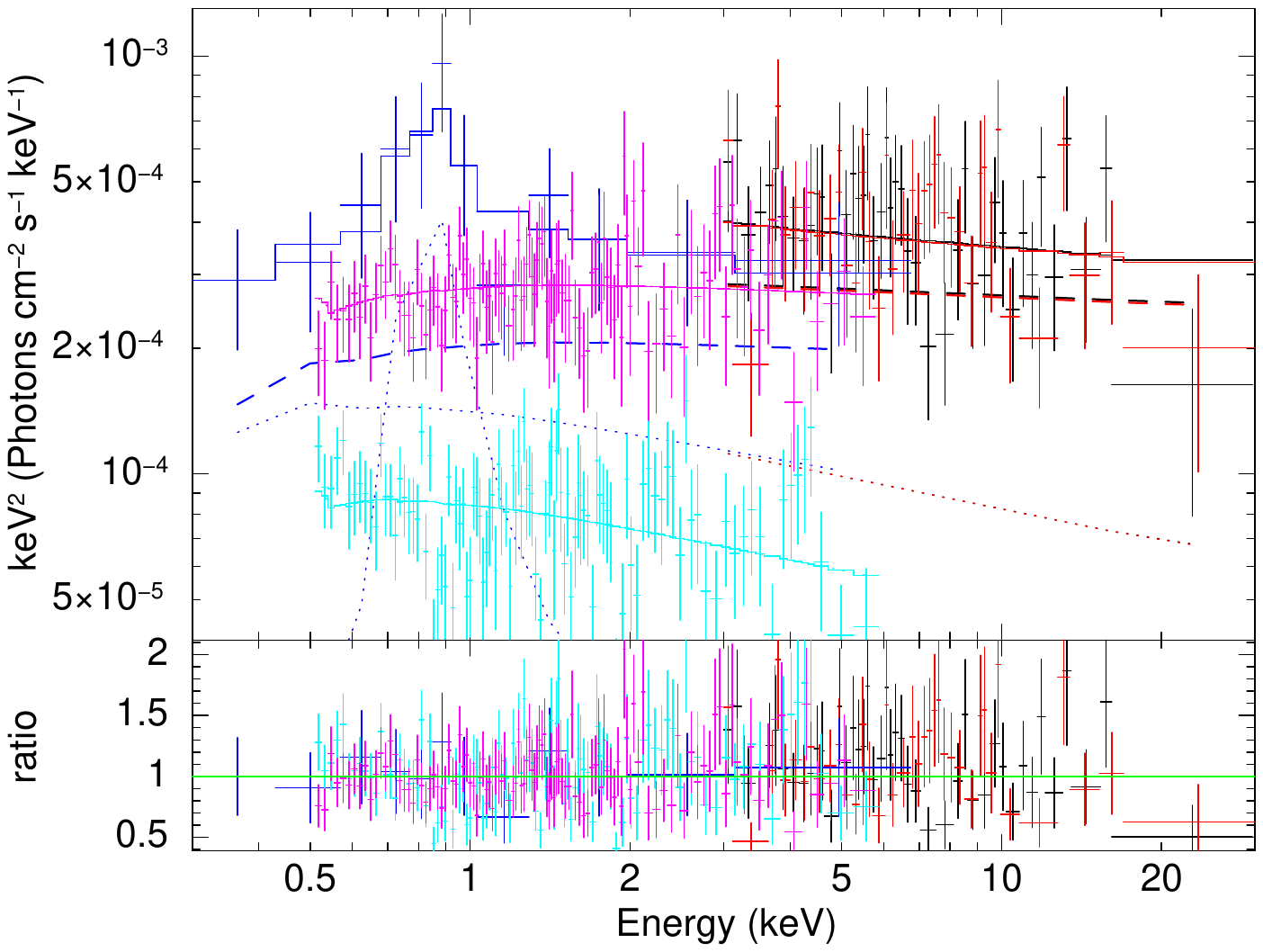}
\includegraphics[width=0.45\textwidth, clip, trim={0.3cm 0.2cm 2.8cm 2.8cm}]{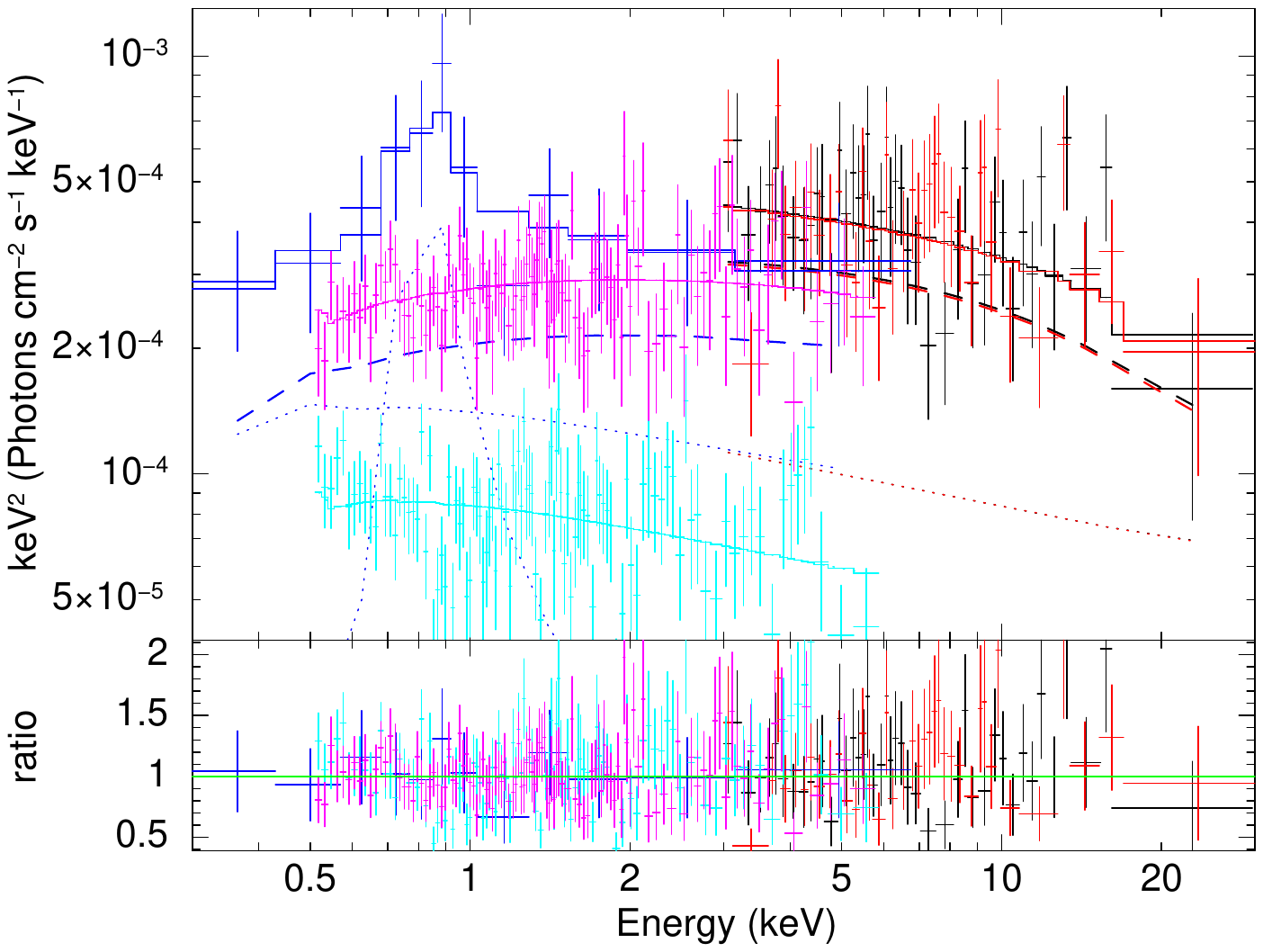}
\caption{X-ray spectra with different models as listed in Table~\ref{table:fit}.  {\tt POWERLAW} models were used on the left-hand panels while {\tt CUTOFFPL} models were included on the corresponding right-hand panels.
Upper left-hand panel: The absorbed {\tt POWERLAW} model of the {\it NuSTAR} spectra (black and red), taken from within a 30\arcsec\ circular region centered on the X-ray peak, was jointly fitted with the {\it Swift} spectrum (blue).  This approach was used to constrain the overall hard X-ray emission originating from AGN activities (Core+Jet).  The dotted lines represent the individual components ({\tt POWERLAW}+{\tt APEC}) as discussed in the text.
Upper right-hand panel: Similar to the upper left-hand panel, but with the {\tt CUTOFFPL} model replacing the {\tt POWERLAW} model.
Middle left-hand panel: Similar to the upper left-hand panel, but including the {\it Chandra} jet spectrum (cyan) modeled with {\tt POWERLAW}.  The flux corrected jet models for the {\it NuSTAR} and {\it Swift} spectra are shown as dotted lines.  The unresolved core is also modeled as a {\tt POWERLAW} (dashed lines).
Middle right-hand panel: Similar to the middle left-hand panel, but with the {\tt CUTOFFPL} model replacing the {\tt POWERLAW} model for the unresolved core.
Lower left-hand panel: Similar to the middle left-hand panel, but including the {\it Chandra} core spectrum (magenta) modeled with {\tt POWERLAW}.
Middle right-hand panel: Similar to the lower left-hand panel, but with the {\tt CUTOFFPL} model replacing the {\tt POWERLAW} model for the unresolved core.
}
\label{fig:spec}
\end{figure*}

\begin{deluxetable*}{ccccccccc}
\tabletypesize{\scriptsize}
\tablewidth{8.5cm}
\tablecolumns{9}
\tablecaption{Best-fit Spectral Results using Phenomenological Models
\label{table:fit}
}
\tablehead{
\colhead{Component} &
\colhead{Parameter} &
\colhead{Unit} &
\multicolumn{2}{c}{Overall$^{\rm a}$} &
\multicolumn{2}{c}{Decomposition} &
\multicolumn{2}{c}{Decomposition}
}
\startdata
\\
Data & & & \multicolumn{2}{c}{{\it NuSTAR}+{\it Swift}} & \multicolumn{2}{c}{{\it NuSTAR}+{\it Swift}} & \multicolumn{2}{c}{{\it NuSTAR}+{\it Swift}} \\
 & & & \multicolumn{2}{c}{} & \multicolumn{2}{c}{+{\it Chandra}(Jet)} & \multicolumn{2}{c}{+{\it Chandra}(Jet+Core)} \\[2mm]
\hline
\\
Core$^{\rm a}$ & Model & & \tt POWERLAW & \tt CUTOFFPL & \tt POWERLAW & \tt CUTOFFPL & \tt POWERLAW & \tt CUTOFFPL \\[2mm]
 & {\it Chandra}$^{\rm b}$ & & no & no & no & no & yes & yes \\[2mm]
 & $\Gamma^{\rm c}$ & \nodata & $2.26^{+0.07}_{-0.07}$ & $1.93^{+0.18}_{-0.19}$ & $2.27^{+0.10}_{-0.10}$ & $1.68^{+0.31}_{-0.35}$ & $2.05^{+0.04}_{-0.04}$ & $1.93^{+0.06}_{-0.06}$ \\[2mm]
 & $E_{\rm cutoff}^{\rm d}$ & keV & \nodata & $20^{+21}_{-7}$  & \nodata & $12^{+12}_{-4}$ & \nodata & $19^{+10}_{-5}$ \\[2mm]
 & norm({\it NuSTAR} A)$^{\rm e}$ & $10^{-4}$\,units$^{\rm k}$ & $5.97^{+0.91}_{-0.79}$ & $4.57^{+0.99}_{-0.84}$ & $4.52^{+0.97}_{-0.80}$ & $2.73^{+1.04}_{-0.83}$ & $3.04^{+0.32}_{-0.30}$ & $3.48^{+0.40}_{-0.36}$ \\[2mm]
 & norm({\it NuSTAR} B)$^{\rm e}$ & $10^{-4}$\,units$^{\rm k}$ & $5.89^{+0.89}_{-0.78}$ & $4.53^{+0.98}_{-0.83}$ & $4.45^{+0.94}_{-0.78}$ & $2.70^{+1.01}_{-0.82}$ & $3.00^{+0.33}_{-0.31}$ & $3.43^{+0.39}_{-0.36}$ \\[2mm]
 & norm({\it Swift})$^{\rm e}$ & $10^{-4}$\,units$^{\rm k}$ & $3.77^{+0.58}_{-0.54}$ & $3.77^{+0.59}_{-0.56}$ & $2.27^{+0.58}_{-0.54}$ & $2.13^{+0.62}_{-0.58}$ & $2.16^{+0.55}_{-0.52}$ & $2.29^{+0.58}_{-0.55}$ \\[2mm]
 & norm({\it Chandra})$^{\rm e}$ & $10^{-4}$\,units$^{\rm k}$ & \nodata & \nodata & \nodata & \nodata & $2.96^{+0.07}_{-0.07}$ & $3.10^{+0.09}_{-0.09}$ \\[2mm]
 & ${f_{\rm 20-30\,keV}}^{\rm f}$ & $10^{-13}$\,erg\,cm$^{-2}$\,s$^{-1}$ & $1.69^{+0.13}_{-0.13}$ & $1.13^{+0.21}_{-0.18}$ & $1.25^{+0.14}_{-0.13}$ & $0.66^{+0.19}_{-0.16}$ & $1.65^{+0.10}_{-0.10}$ & $0.81^{+0.17}_{-0.15}$ \\[2mm]
\hline\\
Jet & Model & & \nodata & \nodata & \tt POWERLAW & \tt POWERLAW & \tt POWERLAW & \tt POWERLAW \\[2mm]
 & {\it Chandra}$^{\rm b}$ & & \nodata & \nodata & yes & yes & yes & yes \\[2mm]
 & $\Gamma^{\rm c}$ & \nodata & \nodata & \nodata & $2.25^{+0.04}_{-0.04}$ & $2.26^{+0.04}_{-0.04}$ & $2.26^{+0.04}_{-0.04}$ & $2.25^{+0.04}_{-0.04}$ \\[2mm]
 & norm$^{\rm e}$ & $10^{-4}$\,units$^{\rm k}$ & \nodata & \nodata & $1.50^{+0.04}_{-0.04}$ & $1.50^{+0.04}_{-0.04}$ & $1.50^{+0.04}_{-0.04}$ & $1.50^{+0.04}_{-0.04}$ \\[2mm]
 & ${f_{\rm 20-30\,keV}}^{\rm f}$ & $10^{-13}$\,erg\,cm$^{-2}$\,s$^{-1}$ & \nodata & \nodata & $0.44^{+0.05}_{-0.04}$ & $0.43^{+0.05}_{-0.04}$ & $0.42^{+0.04}_{-0.04}$ & $0.43^{+0.05}_{-0.04}$ \\[2mm]
\hline\\
 Soft Excess & Model & & \tt APEC & \tt APEC & \tt APEC & \tt APEC & \tt APEC & \tt APEC \\[2mm]
 & $kT^{\rm g}$ & keV & $0.76^{+0.19}_{-0.23}$ & $0.70^{+0.18}_{-0.37}$ & $0.76^{+0.19}_{-0.23}$ & $0.71^{+0.18}_{-0.26}$ & $0.75^{+0.19}_{-0.23}$ & $0.72^{+0.18}_{-0.24}$ \\[2mm]
 & norm$^{\rm e}$ & $10^{-4}$\,units$^{\rm l}$ & $1.09^{+0.56}_{-0.51}$ & $1.27^{+0.56}_{-0.52}$ & $1.09^{+0.56}_{-0.51}$ & $1.34^{+0.57}_{-0.53}$ & $1.20^{+0.56}_{-0.51}$ & $1.18^{+0.54}_{-0.49}$ \\[2mm]
 & ${f_{\rm 0.5-2\,keV}}^{\rm f}$ & $10^{-13}$\,erg\,cm$^{-2}$\,s$^{-1}$ & $2.58^{+1.28}_{-1.20}$ & $3.02^{+1.32}_{-1.24}$ & $2.58^{+1.46}_{-1.04}$ & $3.18^{+1.34}_{-1.26}$ & $2.77^{+1.26}_{-1.18}$ & $2.82^{+1.25}_{-1.17}$ \\[2mm]
\hline\\
Fit Statistics & $C^{\rm h}$ & \nodata & 615.6 & 611.5 & 894.0 & 889.4 & 1108.6 & 1099.0 \\[2mm]
 & d.o.f.$^{\rm i}$ & \nodata & 654 & 653 & 902 & 901 & 1164 & 1163 \\[2mm]
 & AIC$^{\rm j}$ & \nodata & 627.6 & 625.5 & 910.0 & 907.4 & 1126.6 & 1119.0 \\[2mm]
\enddata
\tablenotetext{\rm a}{For the overall X-ray emission, the jet emission is not constrained, and thus the core component includes emission from both the unresolved {\it Chandra} core and the jets.}
\tablenotetext{\rm b}{Whether the component is constrained with \textit{Chandra} spectrum.}
\tablenotetext{\rm c}{Power-law photon index.}
\tablenotetext{\rm d}{Cutoff energy.}
\tablenotetext{\rm e}{Normalizations for the {\it NuSTAR} FPMA/FPMB, {\it Swift}, and {\it Chandra} data.  For the jet and the soft excess components, the normalizations presented correspond to the {\it NuSTAR} and {\it Swift} data, tied together as discussed in the text.}
\tablenotetext{\rm f}{Average unabsorbed flux between the two \textit{NuSTAR} detectors (20--30\,keV) or from the \textit{Swift} detector (0.5--2\,keV).}
\tablenotetext{\rm g}{Plasma temperature.}
\tablenotetext{\rm h}{C-statistics.}
\tablenotetext{\rm i}{Degrees of freedom.}
\tablenotetext{\rm j}{Akaike information criterion, AIC $= C + 2\times k$ , where $k$ is the number of model parameters.}
\tablenotetext{\rm k}{Each unit is equal to 1\,photon\,keV$^{-1}$\,cm$^{-1}$\,s$^{-1}$ at 1\,keV.}
\tablenotetext{\rm l}{Each unit is equal to $\frac{10^{-14}}{4\pi[D_A(1+z)]^2}\int n_e n_H dV$, where $D_A$ is the angular diameter distance to the source (in cm), $dV$ is the volume element (in cm$^3$), and $n_e$ and $n_H$ are electron and hydrogen densities (in cm$^{-3}$), respectively.}
\end{deluxetable*}

\begin{deluxetable*}{cccccc}
\tabletypesize{\scriptsize}
\tablewidth{8.5cm}
\tablecolumns{5}
\tablecaption{Best-fit Spectral Results using Comptonization ({\tt compPS}) Model
\label{table:compps}
}
\tablehead{
\colhead{Component} &
\colhead{Parameter} &
\colhead{Unit} &
\colhead{Overall$^{\rm a}$} &
\colhead{Decomposition} &
\colhead{Decomposition}
}
\startdata
\\
Data & & & {\it NuSTAR}+{\it Swift} & {\it NuSTAR}+{\it Swift} & {\it NuSTAR}+{\it Swift} \\
 & & & & +{\it Chandra}(Jet) & +{\it Chandra}(Jet+Core) \\[2mm]
\hline
\\
Core$^{\rm a}$ & Model & & \tt compPS & \tt compPS & \tt compPS  \\[2mm]
 & {\it Chandra}$^{\rm b}$ & & no & no & yes  \\[2mm]
 & $T_e^{\rm c}$ & keV & $15.3^{+3.2}_{-1.7}$ & $14.0^{+3.1}_{-1.0}$ & $14.9^{+1.7}_{-1.3}$ \\[2mm]
 & $y^{\rm d}$  & \nodata & $0.56^{+0.77}_{-0.09}$ & $0.90^{+4.98}_{-0.41}$ & $0.61^{+0.15}_{-0.06}$ \\[2mm]
 & $\tau^{\rm e}$ & \nodata & 3--12 & 4--58  & 4--7 \\[2mm]
 & norm({\it NuSTAR} A)$^{\rm f}$ & $10^6$\,units$^{\rm l}$ & $5.8^{+8.5}_{-3.6}$ & $1.7^{+3.6}_{-0.5}$ & $3.1^{+1.2}_{-0.9}$ \\[2mm]
 & norm({\it NuSTAR} B)$^{\rm f}$ & $10^6$\,units$^{\rm l}$ & $5.7^{+8.4}_{-3.7}$ & $1.7^{+8.1}_{-0.5}$ & $3.0^{+1.2}_{-0.9}$ \\[2mm]
 & norm({\it Swift})$^{\rm f}$ & $10^6$\,units$^{\rm l}$ & $4.5^{+5.2}_{-2.7}$ & $1.2^{+2.6}_{-0.4}$ & $2.0^{+1.0}_{-0.7}$ \\[2mm]
 & norm({\it Chandra})$^{\rm f}$ & $10^6$\,units$^{\rm l}$ & \nodata & \nodata & $2.7^{+1.0}_{-0.8}$ \\[2mm]
\hline\\
Jet & Model & & \nodata & \tt POWERLAW & \tt POWERLAW \\[2mm]
 & {\it Chandra}$^{\rm b}$ & & \nodata & yes & yes \\[2mm]
 & $\Gamma^{\rm g}$ & \nodata & \nodata & $2.25^{+0.04}_{-0.04}$ & $2.25^{+0.04}_{-0.04}$ \\[2mm]
 & norm$^{\rm f}$ & $10^{-4}$\,units$^{\rm m}$ & \nodata & $1.50^{+0.04}_{-0.04}$ & $1.50^{+0.04}_{-0.04}$ \\[2mm]
\hline\\
 Soft Excess & Model & & \tt APEC & \tt APEC & \tt APEC \\[2mm]
 & $kT^{\rm h}$ & keV & $0.71^{+0.19}_{-0.29}$ & $0.71^{+0.18}_{-0.26}$ & $0.72^{+0.18}_{-0.24}$ \\[2mm]
 & norm$^{\rm f}$ & $10^{-4}$\,units$^{\rm n}$ & $1.21^{+0.55}_{-0.52}$ & $1.22^{+0.54}_{-0.52}$ & $1.17^{+0.54}_{-0.49}$ \\[2mm]
\hline\\
Fit Statistics & $C^{\rm i}$ & \nodata & 612.3 & 890.7 & 1100.2 \\[2mm]
 & d.o.f.$^{\rm j}$ & \nodata & 653 & 901 & 1163 \\[2mm]
 & AIC$^{\rm k}$ & \nodata & 626.3 & 908.7 & 1120.2 \\[2mm]
\enddata
\tablenotetext{\rm a}{For the overall X-ray emission, the jet emission is not constrained, and thus the core component includes emission from both the unresolved {\it Chandra} core and the jets.}
\tablenotetext{\rm b}{Whether the component is constrained with \textit{Chandra} spectrum.}
\tablenotetext{\rm c}{Electron temperature.}
\tablenotetext{\rm d}{Compton parameter $y=4\tau (kT_e/m_e c^2)$.}
\tablenotetext{\rm e}{Optical depth range.}
\tablenotetext{\rm f}{Normalizations for the {\it NuSTAR} FPMA/FPMB, {\it Swift}, and {\it Chandra} data.  For the jet and the soft excess components, the normalizations presented correspond to the {\it NuSTAR} and {\it Swift} data, tied together as discussed in the text.}
\tablenotetext{\rm g}{Power-law photon index.}

\tablenotetext{\rm h}{Plasma temperature.}
\tablenotetext{\rm i}{C-statistics.}
\tablenotetext{\rm j}{Degrees of freedom.}
\tablenotetext{\rm k}{Akaike information criterion, AIC $= C + 2\times k$ , where $k$ is the number of model parameters.}
\tablenotetext{\rm l}{Each unit is equal to $(\frac{R_{\rm in}}{D_{10}})^2 \cos\theta$, where $R_{\rm in}$ is the apparent inner disk radius (in km), $D_{10}$ is the distance to the source (in 10\,kpc), and $\theta$ is the angle of the disk ($\theta=0$ is face-on).}
\tablenotetext{\rm m}{Each unit is equal to 1\,photon\,keV$^{-1}$\,cm$^{-1}$\,s$^{-1}$ at 1\,keV.}
\tablenotetext{\rm n}{Each unit is equal to $\frac{10^{-14}}{4\pi[D_A(1+z)]^2}\int n_e n_H dV$, where $D_A$ is the angular diameter distance to the source (in cm), $dV$ is the volume element (in cm$^3$), and $n_e$ and $n_H$ are electron and hydrogen densities (in cm$^{-3}$), respectively.}
\end{deluxetable*}

We extracted {\it NuSTAR} spectra from the central region of 3C~264  for the observation and created the corresponding response files for the unresolved source (Figure~\ref{fig:spec}, and see the discussion below for further details).
The extraction region is 
circular with a 30\arcsec\ radius, which is close to the  29\arcsec\ 
half-power radius,  centered at the {\it NuSTAR} 3--7\,keV peak 
(Figure~\ref{fig:image}). The background circular region with the same 
radius was chosen to be far enough from the center and also located on 
the same detector chip as the source.  The latter criterion is important 
to minimize instrumental background variations from chip to chip 
\citep{Wik2014a}. During the spectral analysis with {\tt XSPEC}, the two {\it NuSTAR} spectra from both detectors were joint-fitted during the spectral analysis with {\tt XSPEC}.  To constrain the softer emission of the spectrum, we also included the {\it Swift} data taken during the {\it NuSTAR} observation (Section~\ref{sec:obs}).  

The unresolved {\it NuSTAR} X-ray emission is overwhelmingly dominated by the AGN activities, specifically the X-ray core and the jet, with only minimal contributions from the diffuse emission from hot gas in the galaxy, the ICM, or low-mass X-ray binaries (LMXBs).

To detect fainter sources unrelated to the AGN activities, we incorporated a
{\it Chandra} observation taken in 2004 (Section~\ref{sec:obs}; see \citet{Wong2014,Wong2017} for the procedure to detect and analyze {\it Chandra} point sources).  Only one source is detected in the {\it NuSTAR} extraction region with a luminosity of $\sim$$10^{40}$\,erg\,s$^{-1}$ in 0.5--10\,keV if it is at the distance of 3C~264, which is more than two orders of magnitudes fainter than the emission from 3C~264.  \citet{Sun2002} estimated that the contribution from unresolved low-mass X-ray binaries (LMXBs) is at most 4\% using the scaling relation from the LMXB X-ray–to–optical ratios in \citet{Sarazin2001}.  The corresponding luminosity is approximately $1.6\times10^{41}$\,erg\,s$^{-1}$ in 0.5--10\,keV.  To assess the impact of all these point sources, we conservatively incorporated a LMXB model with the aforementioned luminosity \citep{Wong2017,Wik2014b}. The changes in the best-fit photon index and cutoff energy of the LLAGN are negligible, being at most 0.05\% and 0.4\%, respectively.  Thus, we ignore the contributions from these point sources in the spectral analysis.

\subsection{Soft Excess in the {\it Swift} Spectrum}
\label{sec:spec:softexcess}

In the {\it Swift} spectrum, there is an evident excess of X-ray emission around $\sim$0.7--1\,keV beyond what a power-law model predicts. This excess can be described by an {\tt APEC} model absorbed by the Galactic absorption. We fixed the redshift to that of 3C~264 and set the metallicity to 0.8 solar, following the value used by \citet{Sun2007} when fitting the X-ray spectrum of the hot gas of the host galaxy.  The best-fit temperature is approximately 0.7--0.8\,keV.  
The unabsorbed X-ray flux between 0.5 and 2\,keV is about $3 \times 10^{-13}$\,erg\,cm$^{-2}$\,s$^{-1}$ (see Table~\ref{table:fit} below). Assuming the distance to 3C~264, the corresponding X-ray luminosity is about $3 \times 10^{41}$\,erg\,s$^{-1}$.
To determine the significance of this excess and ascertain the necessity of the extra {\tt APEC} model, we simulated 1000 spectra using the single {\tt POWERLAW} model and then compared the likelihood ratio for the single {\tt POWERLAW} model ralative to the {\tt APEC+POWERLAW} model (likelihood ratio test\, {\tt lrt,} in {\tt XSPEC}).  We found that 957 of the simulated likelihood ratios are smaller than the observed ratio ($p$-value = 4.3\%), strongly suggesting that the {\tt APEC+POWERLAW} model is preferred.

However, such a pronounced excess was not observed in the {\it Chandra} data, which was taken 15 yr prior to the {\it NuSTAR} and {\it Swift} observations.  Although \citet{Sun2007} identified hot gas in the host galaxy with a temperature roughly around 0.65\,keV from the {\it Chandra} observation, the flux of the excess in the {\it Swift} data is 20 times higher than the hot gas flux they reported.  Thus, it is unlikely that the soft excess detected with {\it Swift} originates from the diffuse hot gas of the host galaxy.  
Instead, it might be attributed to an increase in soft flux related to AGN activities, such as an additional Comptonization component \citep{Dewangan2007}, ionized reflection \citep{Crummy2006}, or complex absorption \citep{Gierlinski2004,Done2007}.  Alternatively, it could also be attributed to another unresolved transient event, such as an off-nuclear hyperluminous X-ray source (HLX) occurring during the {\it NuSTAR} and {\it Swift} observations.
Fortunately, this excess is soft and sufficiently faint in comparison to the LLAGN and jet spectra, ensuring that it does not significantly influence the spectral fitting results.  These results are robustly anchored by the broader energy range of the spectra.  When excluding the {\tt APEC} component from the fitting, the deviations in the best-fit photon index and cutoff energy are less than the statistical uncertainties.  Nevertheless, the soft excess is taken into account by the absorbed {\tt APEC} model, with both temperature and normalization being free parameters tied across all spectra.

\subsection{Overall Hard X-ray Emission from AGN Activities}
\label{sec:spec:overall}

To constrain the overall hard X-ray emission from the AGN activities, 
which includes the unresolved X-ray core and the jets, we fitted 
the two 3--30\,keV {\it NuSTAR} spectra jointly with the 0.3--10\,keV {\it Swift} spectra.  
All the spectra were grouped with a minimum of one
count per bin and were fitted using the $C$-statistic in {\tt XSPEC}.
Errors of spectral parameters were determined by using $\Delta C 
= 1$ (68\% confidence) for one parameter of interest.

After accounting for the soft excess (Section~\ref{sec:spec:softexcess}), we modeled the {\it Swift} and {\it NuSTAR} spectra using a single {\tt POWERLAW} model to represent the combined emission from both the core and the jet.  To investigate the potential presence of a spectral curvature, which could suggest emission from a hot accretion flow or an X-ray corona proximate to the black hole, we also employed a cutoff power-law ({\tt CUTOFFPL}) model for the fit.  No intrinsic absorption was found in the spectra.  Thus, the same Galactic absorption model was applied to all components, which was fixed at the Galactic value of $N_H = 2.45\times 
10^{20}$~cm$^{-2}$ (COLDEN program from the {\it Chandra} X-ray Center using the data from \citet{Dickey1990}; see also, \citet{Evans2006} and \citet{Perlman2010}).  All the parameters were tied together across all the spectra, with the exception of the normalizations for each spectrum, to account for the cross-calibration uncertainties between the two {\it NuSTAR} and the {\it Swift} detectors.
The best-fit parameters for the overall AGN emission as well as the core and jet components (Section~\ref{sec:decomp} below) are listed in Table~\ref{table:fit}.

The overall spectrum, including the unresolved {\it Chandra} core and the jet, can be sufficiently characterized by a single {\tt POWERLAW} model with a photon index of $\Gamma=2.26^{+0.07}_{-0.07}$, absorbed by the Galactic column density (upper left-hand panel in Figure~\ref{fig:spec}).   
This is consistent with the value of $2.19^{+0.03}_{-0.03}$ determined using the hardness ratio method described in Section \ref{sec:image:profiles}.
This is also consistent with the value of $2.24\pm0.05$ measured by \citet{Perlman2010}.
The overall unabsorbed 20--30\,keV flux, which is completely dominated by the 
central regime of the AGN
activities (core and jet), is $1.69^{+0.13}_{-0.13}\times 10^{-13}$\,erg\,cm$^{-2}$\,s$^{-1}$.  This translates into a luminosity of $1.81^{+0.14}_{-0.13}\times 10^{41}$\,erg\,s$^{-1}$.

For the absorbed {\tt CUTOFFPL} model, the best-fit photon index is $\Gamma=1.93^{+0.18}_{-0.19}$ (upper right-hand panel in Figure~\ref{fig:spec}).  The cutoff energy is $E_{\rm cutoff} = 20^{+21}_{-7}$\,keV with a 3$\sigma$ lower limit of 6\,keV.
The unabsorbed 20--30\,keV flux of the overall AGN-related activities is $1.13^{+0.21}_{-0.18}\times 10^{-13}$\,erg\,cm$^{-2}$\,s$^{-1}$, giving a luminosity of $1.21^{+0.22}_{-0.20}\times 10^{41}$\,erg\,s$^{-1}$.

In the soft X-ray band of 0.5--10\,keV measured with {\it Swift} in 2019, both the {\tt POWERLAW} and {\tt CUTOFFPL} models estimate a luminosity of about $2\times 10^{42}$\,erg\,s$^{-1}$. This is similar to the luminosity of the unresolved core, $1.6\times 10^{42}$\,erg\,s$^{-1}$, observed with {\it Chandra} in 2004 \citep{Perlman2010}, and the power-law luminosity (core) of $2.26\times 10^{42}$\,erg\,s$^{-1}$ observed with {\it XMM-Newton} in 2001 \citep{Donato2004}.  These fluxes are about three to four times lower than the flux observed during a 2018 outburst observed with {\it Chandra} HRC-I \citep{Archer2020}, and it aligns with its low state observed in 2004 and in the preceding years \citep{Boccardi2019}.  These observations suggest that the 3C~264 AGN, as observed by {\it NuSTAR} and {\it Swift} in 2019, is in a quiescent phase.

\subsubsection{Supporting Evidence of a Cutoff in Spectrum}
\label{sec:overall:cutoffevidence}

The {\tt CUTOFFPL} model offers a slightly better fit based on the face
values of the $C$-statistics, but this improvement is not statistically   
significant.  The likelihood ratio test 
({\tt lrt} in {\tt XSPEC})
also does not suggest a
statistical preference for the {\tt CUTOFFPL} model.  
We have calculated the Akaike information criterion \citep[AIC:][]{Akaike1974} to distinguish between models.  The preferred model for the data is the one that minimizes the AIC \citep{Liddle2007, Medvedev2021}. Table~\ref{table:fit} shows that the {\tt CUTOFFPL} model has a slightly smaller AIC than the {\tt POWERLAW} model, suggesting that the {\tt CUTOFFPL} model is slightly preferred.
There is also additional evidence supporting the {\tt CUTOFFPL} model, which is detailed below.

First, a significant discrepancy is observed between the normalizations of
the {\tt POWERLAW} model for {\it Swift} and {\it NuSTAR}, despite the
{\it Swift} observation being conducted simultaneously with the {\it NuSTAR}
observation. No time variation was detected in the {\it NuSTAR} light
curve, suggesting that this discrepancy is not attributable to
variability.
The discrepancy between model normalizations, as measured with {\it
NuSTAR} and {\it Swift}, is approximately 20\% for the {\tt CUTOFFPL} 
model and 45\% for the {\tt POWERLAW} model, respectively. While
a cross-calibration uncertainty of 10--15\% between different X-ray
observatories is within expected limits, and a 20\% difference is on the
high end but not alarming \citep{Madsen2017,Molina2019,Abdelmaguid2023}, a discrepancy exceeding 40\% is inconsistent 
with the current calibration standards\footnote{See the current
calibration status presented at the International Astronomical Consortium
for High-Energy Calibration: \url{https://iachec.org/}}.
The discrepancy observed in the single {\tt POWERLAW} model could
potentially be explained by adjusting the normalizations to account for
spectral curvature.

More specifically, the fitting is biased toward the {\it NuSTAR} data
because there are more photon counts in the {\it NuSTAR} data.  When fitting a
single {\tt POWERLAW} model to both the {\it NuSTAR} and {\it Swift}   
datasets and tying their photon indices, if there is a cutoff in energy,
the {\tt POWERLAW} slope will try to fit the steeper {\it NuSTAR} spectra 
at higher energy.  Therefore, if the normalization of {\it
Swift} is the same as the {\it NuSTAR} normalization, the model will
overestimate the softer photons measured by {\it Swift}.  To compensate
for this curvature in the fitting, the {\it Swift} normalization needs to
be lower than the {\it NuSTAR} normalization.

We simulated the overall spectra with the absorbed {\tt CUTOFFPL}+{\tt
APEC} model using 1000 times longer exposures for each detector.  The 
model parameters used are listed in Table~\ref{table:fit}, except that we set
all the normalizations of the {\tt CUTOFFPL} model to those determined by
the {\it NuSTAR} FPMA detector.  The best-fit models to the simulated
datasets largely reproduce this normalization discrepancy, with the {\it
NuSTAR} normalizations being more than 30\% higher than that of {\it
Swift}\footnote{If we increase the exposure time of {\it Swift} 
relative to {\it NuSTAR} in the simulation (e.g., simulating all
observations with an exposure of 100\,ks), {\it Swift} will have many more
counts and the single {\tt POWERLAW} will follow the {\it Swift} data. 
The discrepancy between the {\it NuSTAR} and {\it Swift} normalizations is
no longer noticeable.}.

\begin{figure}
\includegraphics[width=0.5\textwidth]{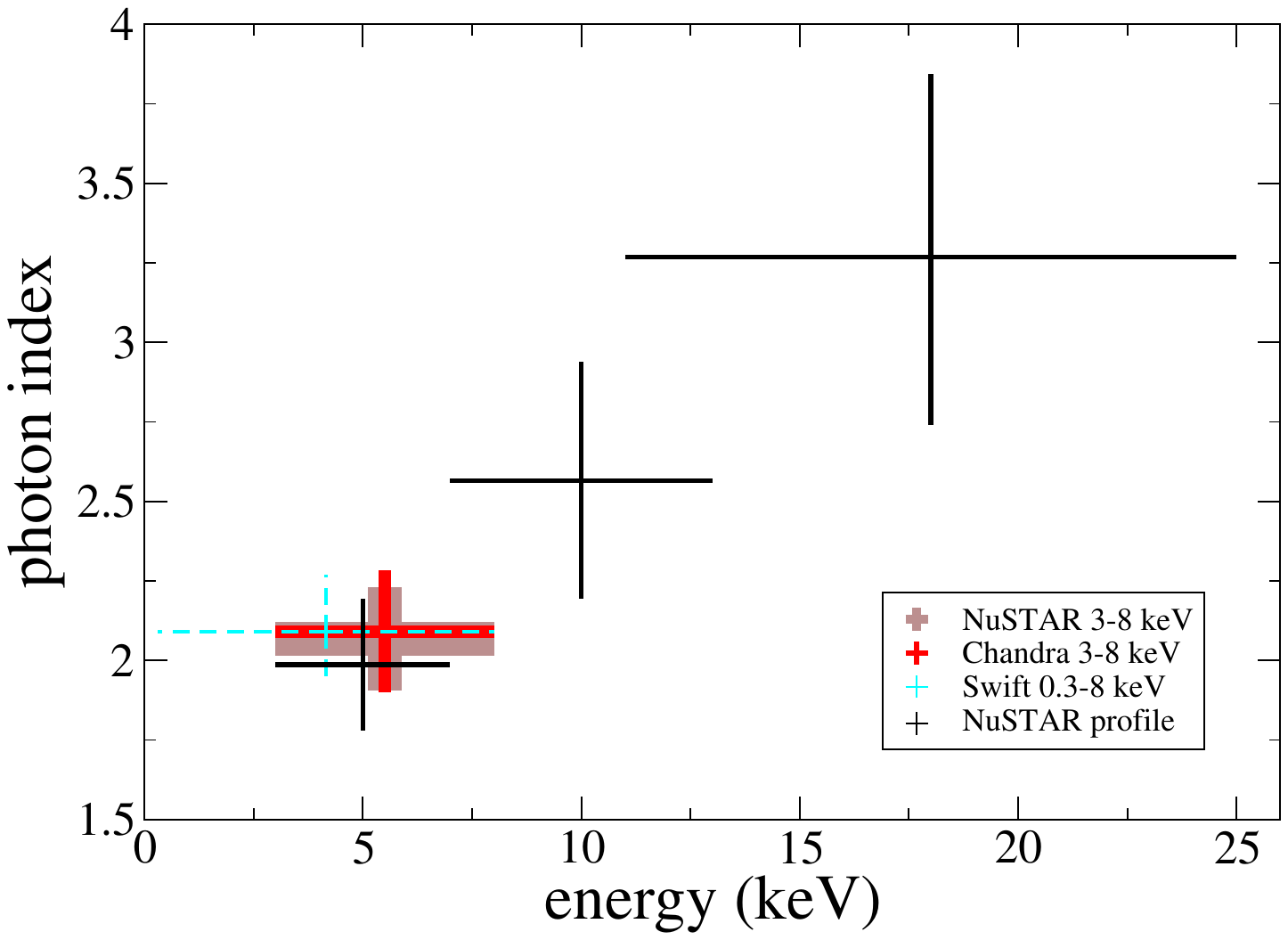}
\caption{Photon index fitted at various energy bins for different dataset.  {\it NuSTAR} and {\it Chandra} data were fitted with absorbed {\tt POWERLAW} model.  For {\it Swift} data, since there is a clear soft excess, an absorbed {\tt APEC + POWERLAW} model was used.
}
\label{fig:pho_vs_E}
\end{figure}

Second, we divided the {\it NuSTAR} spectra into different energy bands and fit the {\it NuSTAR} data separately for each band using a {\tt POWERLAW} model. For comparison, we also fitted the {\it Chandra} and {\it Swift} data individually. The spectral slope steepens (i.e., the photon index increases) at higher energies (see Figure~\ref{fig:pho_vs_E}), indicating the presence of an energy cutoff.

Third, when decomposing the core and jet spectra with {\it Chandra} data (Section~\ref{sec:decomp:jetcore}), the {\tt CUTOFFPL} model is statistically favored for the core emission, suggesting that a cutoff is needed at least for the core.

These points serve as supporting evidence for the {\tt CUTOFFPL} model. However, to definitively confirm the presence of an energy cutoff, deeper simultaneous observations spanning a broad energy range (e.g., 0.5--40\,keV) would be necessary.

\subsubsection{Comptonization Model with Reflection}
\label{sec:overall:compps}

We replace the phenomenological {\tt POWERLAW} or the {\tt CUTOFFPL} models with the {\tt compPS} model \citep{Poutanen1996}, which is a physical model that includes the thermal Comptonization emission from hot plasma and reflection from a standard accretion disk.  Following \citet{Ursini2015}, we set the geometry of the hot plasma to spherical (${\tt GEOM}=-4$). The disk is assumed to have a multicolor temperature, with the inner disk temperature set to 10\,eV
\citep{Beckmann2011, Ursini2015} and the inclination angle set to $i=50\arcdeg$ \citep{Donato2004}. The free parameters for the {\tt compPS} model include the hot plasma temperature $T_e$, the Compton parameter $y$, reflection strength $R$, and normalizations.

While the reflection strength cannot be precisely constrained, the best-fit value is effectively zero ($R=4\times 10^{-14}$), and the 1-$\sigma$ upper limit is $R \le 0.68$. Therefore, since there is no evidence of a reflection hump, we fixed $R=0$ in the fitting process.

The best-fit parameters for the overall AGN emission as well as the core and jet components (Section~\ref{sec:decomp} below) are listed in Table~\ref{table:compps}.  The best-fit hot plasma temperature is $T_e = 15^{+3}_{-2}$\,keV, and the Compton parameter is $y=0.56^{+0.77}_{-0.09}$. The optical depth, $\tau$, is related to the Compton parameter and temperature by $y=4\tau (kT_e/m_e c^2)$.  Therefore, the optical depth falls within the range of $\tau = 3$--$12$.  Using different geometrical settings for the hot plasma (e.g., slab, cylinder, or hemisphere) gives similar results.

\subsection{Decomposing the Hard X-ray Emission of the Core and Jet}
\label{sec:decomp}

\subsubsection{With {\it Chandra} Jet Spectrum}     
\label{sec:decomp:jet}

We further constrained the X-ray emission from the individual components
of the unresolved X-ray core and the jet by jointly fitting their
corresponding {\it Chandra} spectra\footnote{Given that the fluxes 
measured by {\it Chandra} and {\it NuSTAR} in their overlapping 3--8\,keV
energy band are consistent with each other to within approximately 9\%, 
which is comparable to the cross-calibration uncertainties, we
can infer that the quiescent spectra of the core and the jet are stable to this
degree of variance.  Notably, the jet's size scale resolved in the {\it   
Chandra} deconvolved image exceeds 0\farcs2, or 88\,pc.  Therefore, it is
reasonable to assume that the jet flux resolved by {\it Chandra} remained
stable over the roughly 10 year span between the {\it Chandra} and {\it
NuSTAR} observations during its quiescent state.  However, it is also known that jet knots can
exhibit variability on timescales ranging from months to years
\citep[e.g., HST-1, ][]{Harris2006}.}.  
We first included the {\it Chandra}
jet spectrum, extracted from the partial annular regions between 0\farcs2
and 1\farcs2, as shown in the lower left-hand panel of
Figure~\ref{fig:image_arestore}.  This jet spectrum was modeled as an
extra absorbed {\tt POWERLAW}.  The power-law indices for the {\it
NuSTAR}, {\it Swift}, and {\it Chandra} spectra were tied together.
Given that the {\it Chandra} PSF size scale is comparable to the jet
spectral extraction region, some photon leakage from this region is  
anticipated. However, for the {\it NuSTAR} and {\it Swift} spectra, the
spectral extraction regions are expected to encompass most of the jet
emission.  By counting the total number of photons in the deconvolved  
image within a rectangular region that includes most of the jet's photons
and comparing this count to that observed in the {\it Chandra} jet 
spectrum, we estimated that the jet spectrum accounts for 59.6\% of the   
total counts.  We corrected for this flux loss by tying the {\it NuSTAR}
and {\it Chandra} normalizations to the ratio of this fraction.  The {\it
Swift} and {\it NuSTAR} normalizations were also tied together.  Based on
the deconvolved image, we assessed that the emission beyond 1\farcs2
contributes less than 5\% of the total emission.  However, for the
emission inside 0\farcs2, we lack a precise estimate of the unresolved jet's
contribution to the X-ray emission.  Thus, we did not apply such
corrections in the spectral analysis.

For the {\it NuSTAR} and {\it Swift} spectra, after accounting for the jet
emission, the remaining emission predominantly originates from the
unresolved {\it Chandra} core. Similar to Section~\ref{sec:spec:overall},
we modeled this emission with either an absorbed {\tt POWERLAW} model or 
an absorbed {\tt CUTOFFPL} model.  Additionally, we kept the
normalizations for {\it NuSTAR} and {\it Swift} separate (untied).   

The best-fit spectra with all the model components are displayed in the   
middle panels of Figure~\ref{fig:spec}.  The best-fit photon indices for   
the {\tt POWERLAW} jet component, when jointly fitted with either the
single {\tt POWERLAW} or the {\tt CUTOFFPL} models of the core, are
$2.25^{+0.04}_{-0.04}$ and $2.26^{+0.04}_{-0.04}$, respectively.  For the
core, the single {\tt POWERLAW} model yields a photon index of
$2.27^{+0.10}_{-0.10}$, which is essentially the same as the jet's index.    
The {\tt CUTOFFPL} model for the core has a significantly harder photon
index of $1.68^{+0.31}_{-0.35}$, and the cutoff energy is determined to be
$E_{\rm cutoff} = 12^{+12}_{-4}$\,keV.  The 20--30\,keV flux of the core, as
measured with {\it NuSTAR}, is $1.25^{+0.14}_{-0.13}\times
10^{-13}$\,erg\,cm$^{-2}$\,s$^{-1}$ for the {\tt POWERLAW} model and
$0.66^{+0.19}_{-0.16}\times 10^{-13}$\,erg\,cm$^{-2}$\,s$^{-1}$ for the
{\tt CUTOFFPL} model, accounting for about 74\% and 58\% of the overall   
AGN emission, respectively.

In the case of the {\tt POWERLAW} model for the core component, the 
discrepancy between the {\it NuSTAR} and {\it Swift} normalizations is    
larger compared to the analysis presented in 
Section~\ref{sec:spec:overall}.  The {\it NuSTAR} normalizations are twice
as high as those from {\it Swift}, and they are inconsistent at a level   
beyond 
90\% confidence
statistically.  For the {\tt CUTOFFPL} model, however,   
the {\it NuSTAR} normalizations are about 25\% higher than the {\it Swift}
normalization, and they are consistent within 1$\sigma$ statistically.
This further supports the hypothesis of a necessary cutoff in the
power-law model for the core.

Using the {\tt compPS} model discussed in Section~\ref{sec:overall:compps} for the core, the best-fit hot plasma temperature is determined to be $T_e = 14^{+3}_{-1}$\,keV, and the Compton parameter is $y=0.9^{+5.0}_{-0.4}$. The optical depth is within the range of $\tau = 4$--$58$.

\subsubsection{With {\it Chandra} Jet and Core Spectra}
\label{sec:decomp:jetcore}

Finally, we extracted a {\it Chandra} core spectrum within a radius of    
0\farcs2. PSF correction was applied when generating the response file.   
As discussed in Section~\ref{sec:decomp:jet}, the amount of emission
originating from the unresolved jet within 0\farcs2 is unknown, and
therefore we did not include any corrections for this.

We additionally included the {\it Chandra} core spectrum and jointly
fitted it with the {\it NuSTAR} and {\it Swift} spectra.  We modeled the
core component using either a single {\tt POWERLAW} or a {\tt CUTOFFPL}
model.  Given the potential for variability, the spectral shapes of the
{\it Chandra} and {\it NuSTAR}/{\it Swift} observations might differ.  To
investigate the potential spectral variation, we separately fitted the
{\it Chandra} and {\it NuSTAR} spectra in the overlapping 3--8\,keV energy
band using an absorbed {\tt POWERLAW} model.  For the {\it Swift}
spectrum, we extended the fitting range to 0.3--8\,keV to increase the
statistics and included an {\tt APEC} model, as described in
Section~\ref{sec:spec:softexcess}, to account for the soft excess. The
photon indices from all three instruments are consistent within a
1$\sigma$ level, indicating no evidence of spectral variation.
 Consequently, we tied the spectral indices of the core components across
all observations during the fitting process.  The normalizations for {\it
NuSTAR}, {\it Swift}, and {\it Chandra} were kept separate (untied) to   
account for calibration uncertainties or potential flux variations.

The best-fit spectra with all model components are shown in the lower
panels of Figure~\ref{fig:spec}. The best-fit photon indices of the jet 
component, when jointly fitted with the single {\tt POWERLAW} and the {\tt
CUTOFFPL} core models, are $2.26^{+0.04}_{-0.04}$ and 
$2.25^{+0.04}_{-0.04}$, respectively.  These values are essentially the
same as those obtained from the analysis without the {\it Chandra}
core spectrum.

For the core, the single {\tt POWERLAW} model yields a photon index of    
$2.05^{+0.04}_{-0.04}$, which is significantly harder than the jet’s 
spectrum.  The photon index for the {\tt CUTOFFPL} model is 
$1.93^{+0.06}_{-0.06}$, which is also notably harder.  The cutoff energy is 
determined to be $E_{\rm cutoff} = 19^{+10}_{-5}$\,keV.  The associated 
20--30\,keV flux measured with {\it NuSTAR} is $1.65^{+0.10}_{-0.10}\times  
10^{-13}$\,erg\,cm$^{-2}$\,s$^{-1}$ and $0.81^{+0.17}_{-0.15}\times 
10^{-13}$\,erg\,cm$^{-2}$\,s$^{-1}$ for the {\tt POWERLAW} and {\tt CUTOFFPL}
models, respectively.  These values account for approximately 98\% and      
72\% of the overall AGN emission, respectively.

In the case of both the {\tt POWERLAW} and {\tt CUTOFFPL} models for the  
core component, while the {\it Swift} data normalizations are 40--50\%
lower than those of {\it NuSTAR} or {\it Chandra}, they remain 
statistically consistent at the 90\% confidence level.  Notably, the {\it   
NuSTAR} and {\it Chandra} normalizations are consistent within 1$\sigma$.
This supports the assumption that there is no significant spectral
variation.

We conducted a likelihood ratio test ({\tt lrt})
in {\tt XSPEC} with 1000 simulations
to differentiate between the {\tt POWERLAW} and {\tt CUTOFFPL} models for
the core component.  The test results suggest that the {\tt POWERLAW}
model is rejected at a confidence level of 5.4\%, strongly indicating the 
need for a cutoff in the spectrum.

The {\tt compPS} model for the core, as discussed in section~\ref{sec:overall:compps}, yields a best-fit hot plasma temperature of $T_e = 14.9^{+1.7}_{-1.3}$\,keV and a Compton parameter of $y=0.61^{+0.15}_{-0.06}$. The optical depth falls within the range of $\tau = 4$--$7$.

\section{Discussion}
\label{sec:discussion}

We have detected hard X-ray emission $\gtrsim 10$\,keV from 3C~264.  However, both the soft and hard X-ray emissions are unresolved by {\it NuSTAR}.  The locations of the soft and hard X-ray peaks are consistent with each other, indicating that the origin of the hard X-ray emission aligns with that of the soft emission, likely from the core and the jet resolved by {\it Chandra} and by other optical and radio observations.

We have presented the highest spatial resolution X-ray {\it Chandra} image 
using the Lucy-Richardson deconvolution method. The X-ray jet can be 
distinctly resolved down to approximately 0\farcs2 from the center of the 
unresolved core. The X-ray morphology bears a remarkable similarity to the 
radio and optical emissions.  The X-ray jet is most luminous within about 
0\farcs8 from the core and appears linear, indicating that the jet's 
momentum is dominant.  This region, at or less than 0\farcs8 from the 
core, is also where four knobs have been resolved with {\it HST}, ranging 
from 0\farcs15 to 0\farcs6 away from the core center 
\citep{Perlman2010,Meyer2015}.
Beyond that, the jet significantly diminishes in brightness and 
splits into two directions. The stronger path continues northeast, while 
the weaker one turns to the east. X-ray emissions from the jet are 
undetectable beyond $\sim$2\arcsec.

The deconvolved {\it Chandra} images in the 0.5--1\,keV and 1--6\,keV bands have been utilized to create a hardness ratio map, allowing us to investigate the spectral properties at a spatial resolution surpassing that achievable through spectral analysis. The average photon index of the jet, determined by this hardness ratio map within the soft X-ray emission range (0.5--6\,keV), is $\Gamma=2.26^{+0.05}_{-0.05}$. This finding is in excellent agreement with the spectral analysis, which yields a $\Gamma$ ranging between 2.21 and 2.30 (see Table~\ref{table:fit}). The hardness ratio map also indicates that the photon index of the unresolved {\it Chandra} core (within $< 0\farcs2$) is significantly harder than in the adjacent regions of the jet.

When jointly fitting the {\it NuSTAR} spectra with the {\it Swift} spectrum taken during the {\it NuSTAR} observation, the overall AGN spectrum (core+jet) can be adequately modeled with either the {\tt POWERLAW} or {\tt CUTOFFPL} models.  However, there is evidence supporting the presence of a cutoff energy at $E_{\rm cutoff} = 20^{+21}_{-7}$\,keV.  Including the resolved {\it Chandra} core and jet spectra in the soft X-ray, taken 15 yr prior to the {\it NuSTAR} and {\it Swift} observations, further reinforces the likelihood of a cutoff at $E_{\rm cutoff} = 19^{+10}_{-5}$\,keV.  The {\tt CUTOFFPL} model also indicates a significantly harder photon index for the core, aligning with the hardness ratio analysis.  Such a cutoff in energy implies that the X-ray emission from the core can be at least partially contributed by the RIAF.  Notably, the cutoff energy for 3C~264 is the lowest among other LLAGNs measured in X-ray \citep{Chakraborty2023,Jana2023}. 

When fitted with the Comptonization model ({\tt compPS}), the best-fit electron temperature is approximately $T_e\approx15$\,keV, and the optical depth is more precisely constrained within the range of $\tau=4$--7, especially when combined with {\it Chandra} data.  This aligns excellently with the anticorrelation observed between optical depth and electron temperature in a sample of 16 LLAGNs \citep{Chakraborty2023}.  Such an anticorrelation has also been previously noted in more luminous Seyfert galaxies \citep{Tortosa2018} and in the hard state of black hole binaries \citep{Banerjee2020}, suggesting the universality of coronal properties across different black hole masses and accretion rates.  This has important implications for a departure from a fixed disk-corona configuration in radiative balance \citep[see][for further discussions]{Tortosa2018,Chakraborty2023}.

The overall X-ray spectrum is consistent with the X-ray properties of 
LLAGNs discussed by \citet{Ho2008}.
The photon index between 0.5 and 10~keV is approximately 2.1, which is within the range 
of $1.4$--$2.2$ of typical LLAGNs.
The power-law component shows very little intrinsic absorption.
There is no evidence of Fe K$\alpha$ emission or Compton reflection, suggesting either the absence of an optically thick accretion disk or that it is truncated and replaced by an optically thin RIAF.
Note that RIAF models tend to predict harder X-ray spectra than jet models, as can be seen in Figure 17 of \cite{Nemmen2014}.  As such, the measured photon index of 1.93 from the {\tt CUTOFFPL} core model agrees better with a RIAF origin.
Detailed modeling of the multiwavelength SED will be needed to confirm this \citep[e.g.,][]{Nemmen2014}.

A similar conclusion regarding the existence of RIAFs has been drawn from studies of the LLAGNs Cen~A\footnote{Although evidence of reflection has been reported in earlier work by \citet{Fukazawa2011} and \citet{Burke2014}; see also a discussion by \citet{Beckmann2011}.} \citep{Furst2016}, M81 \citep{Young2018}, NGC~7213 \citep{Ursini2015}, NGC~3998 and NGC~4579 \citep{Younes2019}.
Although Fe lines have been detected in a number of cases, they are attributed to hot thermal plasma or to broad-line regions \citep[see, e.g.,][]{Ursini2015, Young2018, Younes2019}.
For 3C~264, excess in soft emission below $\sim$2\,keV is detected, which can be fitted with a thermal plasma model with a temperature of 0.7--0.8\,keV.  The origin of the soft excess is not clear.
All of these are consistent with the X-ray properties of LLAGNs as described by \citet{Ho2008}.

\subsection{Implication  of the Low Cutoff Energy}
\label{sec:discussion:lowEcut}

The cutoff energy $E_{\rm cutoff}$$\approx$20\,keV of 3C~264 is very low when compared to other AGNs, which have a median  $E_{\rm cutoff}$$\sim$200\,keV \citep{Ricci2017}.  Lower cutoff values are usually observed at very high Eddington ratios \citep{Ricci2018}. 
The LLAGN 3C~264 has an Eddington ratio of approximately $7\times10^{-5}$.  This ratio places it outside the typical correlations observed for AGNs with Eddington ratios higher than about 0.001, such as the photon index-Eddington ratio correlation noted in recent studies \citep[e.g.,][]{Chakraborty2023,Jana2023}.  For instance, these cited recent studies do not report higher cutoff energies for LLAGNs with Eddington ratios below 0.001, unlike those expected with higher ratios.  Some LLAGNs exhibit very low cutoff energies or coronal temperatures, such as a cutoff energy $E_{\rm cutoff}$$\approx$80\,keV and an electron temperature $T_e$$\approx$30--40\,keV in low-accreting Seyferts \citep{Jana2023} and an electron temperature $T_e$$\lesssim$10--20\,keV in LLAGNs \citep{Chakraborty2023}.  Therefore, the notably low cutoff of 3C~264, considering the uncertainties, is not overly surprising.

The low cutoff energy suggests a cooler corona or plasma within the hot accretion flow.  If the X-ray emission arises from the thermal Comptonization of seed disk photons by this hot plasma, particularly in a scenario where the disk is truncated in a RIAF, Comptonization could effectively cool the hot plasma.  The smaller the truncation radius, the more efficient the cooling becomes \citep{Tortosa2018}.  Consequently, the low cutoff energy indicates a small truncation radius.  However, there is no evidence of a truncated disk in 3C~264, such as reflection features or a UV bump  \citep{Boccardi2019}, and the presence of a truncated disk remains unconfirmed.  Alternatively, a reduced scale height of the corona or hot accretion flow could also enhance the Compton cooling rate of the hot plasma \citep{Tortosa2018}.  The cutoff might also result from bremsstrahlung emission in the outer parts of the RIAF near the Bondi radius.  This extended quiescent emission is similarly proposed to explain the low cutoff energy in the X-ray spectrum of Sgr~A$^*$ \citep{Yuan2003,Quataert2002}.

\subsection{Testing the Synchrotron Self-Compton Model}
\label{sec:discussion:SSC}

To account for the $\gamma$-ray and TeV emission from 3C~264, the SED has been modeled using the synchrotron self-Compton (SSC) 
jetted model \citep[Figure~1 in][]{Kagaya2017}.
This model adequately fits the softer X-ray emission (up to $\lesssim10$\,keV) and the $\gamma$-ray emission observed during the quiescent state of the core.  However, it significantly underpredicts the hard X-ray emission (around $\sim$20\,keV) observed with {\it NuSTAR} by an order of magnitude; our measured SED at 20\,keV is around $\nu f_{\nu} = 3$--$4\times 10^{-13}$\,erg\,cm$^{-2}$\,s$^{-1}$ while the SSC model predicts a value of $\sim$$4\times 10^{-14}$\,erg\,cm$^{-2}$\,s$^{-1}$.  
Furthermore, the predicted power-law index of approximately 3 in 1--10\,keV is substantially steeper than the value measured.
The large discrepancy is likely due to the phenomenological model used to fit the emission below the hard X-ray band.
Similarly, recent SSC jetted models of the low state of the 3C 264 core also underestimated the SED at $\sim$20\,keV by more than an order of magnitude, and their predicted power-law slope in the 1--10\,keV range of $\sim$3--4 is also significantly steeper than our observed values \citep{Boccardi2019}.
It is possible that these model fittings underestimated the maximum energy of the electron population in 3C~264.
This contrasts with the findings for another TeV radio galaxy, M87, where the SSC model overpredicts the hard X-ray emission at 40\,keV by approximately a factor of three during the quiescent state of the core, and the predicted power-law index is flatter than the measured spectral slope \citep{Wong2017}.  {\it NuSTAR} plays a crucial role in exploring the transition from synchrotron-dominated to (self) inverse Compton-dominated emission around 10\,keV, thereby providing key insights into these VHE processes.  

Presently, our understanding is limited by the statistical uncertainties of our data.  Future in-depth observations with {\it NuSTAR}, extending detection beyond 30\,keV, and simultaneous high angular resolution multiwavelength observations, will offer tighter constraints on accretion models and the VHE emission mechanisms.

\section{Conclusion}
\label{sec:conclusion}

For the first time, hard X-ray emission $\gtrsim 10$\,keV from the TeV radio galaxy 3C~264 has been detected, which is unresolved with {\it NuSTAR}.  The location of the hard X-ray emission is consistent with that of the soft X-ray emission.  An excess in soft emission below approximately 2\,keV has been detected with {\it Swift}, and its origin is unclear.  We have generated high-resolution deconvolved {\it Chandra} images to more precisely determine the origin of the X-ray emission. These deconvolved {\it Chandra} images show that the X-ray jet can be resolved down to approximately 0\farcs2 from the unresolved {\it Chandra} core.  The X-ray morphology is similar to that of the radio and optical jet. Furthermore, the X-ray spectrum of the {\it Chandra} core is harder than that of the jet.  The X-ray spectrum can be adequately modeled with a single power-law model. However, evidence suggests the presence of a cutoff in the energy around 20\,keV, which indicates that at least some of the X-ray emission from the core can be attributed to the RIAF. The Comptonization model indicates an electron temperature of about 15\,keV and an optical depth ranging between 4 and 7, following the universality of coronal properties of black hole accretion.  The cutoff energy or electron temperature of 3C 264 is the lowest among those of other LLAGNs measured in X-rays. Meanwhile, recent SSC models, which were proposed to explain $\gamma$-ray and TeV emission, significantly underestimate the hard X-ray emission from the low state of the core at 20\,keV by an order of magnitude or more.
This suggests that the synchrotron electrons might be accelerated to higher energies than previously thought.

\section*{Acknowledgments}
The authors would like to thank the referee and the editors for their comments and suggestions, which have improved the manuscript.
This work was supported by the following NASA grants: \textit{NuSTAR} awards 80NSSC21K1855, 80NSSC20K0052, 80NSSC22K0032, and 80NSSC22K0065, \textit{Chandra} awards GO7-18071B, G09-20111A, and G02-23092X.  C.S. and A.B. were also supported by the SUNY Brockport's Summer Undergraduate Research Program (SURP) and the Physics Department's Richard V. Mancuso Summer Research Award and Donald '80 and Diana '81 Hallenbeck Research Scholars Fund. R.N. was supported by NASA through the \textit{Fermi} Guest Investigator Program (Cycle 16) and a Bolsa de Produtividade from Conselho Nacional de Desenvolvimento Cient\'ifico e Tecnol\'ogico.  D.L. was supported by the NASA ADAP grant 80NSSC22K0218.
This paper employs a list of {\it Chandra} datasets, obtained by the {\it Chandra} X-ray Observatory, which are contained in doi:\href{https://doi.org/10.25574/cdc.263}{10.25574/cdc.263}.

\bibliographystyle{aasjournal}
\bibliography{references}

\end{document}